\documentclass[12pt]{article}

\pdfoutput=1
\usepackage{jheppub}
\usepackage{graphicx,epsfig,amsmath,amssymb}
\usepackage{subcaption}
\usepackage{multirow}
\usepackage{slashed}

\newcommand{\als}{\ensuremath{\alpha_s}}

\newcommand{\amp}{\ensuremath{\mathcal{M}}}

\def\nn{\nonumber}

\def\als{\alpha_s}

\def\eps{\epsilon}

\def\be{\begin{equation}}
\def\ee{\end{equation}}
\def\bea{\begin{eqnarray}}
\def\eea{\end{eqnarray}}
\def\nn{\nonumber}

\def\amp{{\mathcal A}}

\newcommand{\secdec}{\textsc{SecDec}{}}
\newcommand{\pysecdec}{py\secdec}

\newcommand{\mt}{m_t}
\newcommand{\mz}{m_Z}
\newcommand{\mh}{m_H}
\newcommand{\pth}{p_{T,H}}

\title{$ZH$ production in gluon fusion: two-loop amplitudes with full top quark mass dependence}
\author[a,f]{Long Chen,}
\author[b]{Gudrun Heinrich,}
\author[c,d]{Stephen P.~Jones,}
\author[e]{Matthias Kerner,}
\author[f]{Jonas Klappert,}
\author[g]{Johannes Schlenk}

\affiliation[a]{Max Planck Institute for Physics, F\"ohringer Ring 6, 80805 M\"unchen, Germany}
\affiliation[b]{Institute for Theoretical Physics, Karlsruhe Institute of Technology, 76128 Karlsruhe, Germany}
\affiliation[c]{Theoretical Physics Department, CERN, Geneva, Switzerland}
\affiliation[d]{Institute for Particle Physics Phenomenology, Durham University, Durham DH1 3LE, UK}
\affiliation[e]{Physik-Institut, Universit{\"a}t Z{\"u}rich, Winterthurerstrasse 190, 8057 Z{\"u}rich, Switzerland}
\affiliation[f]{Institute for Theoretical Particle Physics and Cosmology, RWTH Aachen University, 52056 Aachen, Germany}
\affiliation[g]{Theory Group LTP, Paul Scherrer Institut, CH-5232 Villigen PSI, Switzerland}

\emailAdd{longchen@physik.rwth-aachen.de}
\emailAdd{gudrun.heinrich@kit.edu}
\emailAdd{s.jones@cern.ch}
\emailAdd{mkerner@physik.uzh.ch}
\emailAdd{klappert@physik.rwth-aachen.de}
\emailAdd{johannes.schlenk@psi.ch}

\preprint{
  {\small 
    \hphantom{.}\hfill ZU-TH 45/20\\
    \hphantom{.}\hfill CERN-TH-2020-199\\
    \hphantom{.}\hfill IPPP/20/57\\
    \hphantom{.}\hfill P3H-20-076\\
    \hphantom{.}\hfill KA-TP-21-2020\\
   \hphantom{.}\hfill TTK-20-42\\
   \hphantom{.}\hfill PSI-PR-20-21}\\
}

\abstract{
We present results for the two-loop helicity amplitudes entering the NLO QCD corrections to the production of a Higgs boson in association with a $Z$-boson in gluon fusion.
The two-loop integrals, involving massive top quarks, are calculated numerically.
Results for the interference of the finite part of the two-loop amplitudes with the Born amplitude are shown as a function of the two kinematic invariants on which the amplitudes depend.
}


\begin{document}

\maketitle

\section{Introduction}

The production of a Higgs boson in association with a vector boson, also called {\em Higgs-Strahlung}, is an important process
at the LHC. For example, this process was used to discover the decay
of Higgs bosons to $b$-quark pairs~\cite{Aaboud:2018zhk,Sirunyan:2018kst},
and it is very well suited to constrain anomalous couplings of the Higgs boson in both the Yukawa and the gauge boson sector.
The production of a Higgs boson in association with a leptonically decaying $Z$-boson facilitates triggering independently of the Higgs boson decay.
This makes this channel especially attractive in combination with challenging Higgs decays, like invisible or hadronic decays, in particular $H\to b\bar{b}$. The regime of boosted Higgs bosons is particularly interesting, as the large-$\pth$ region is sensitive to new physics~\cite{Mimasu:2015nqa,Greljo:2017spw,Alioli:2018ljm,Banerjee:2018bio,Banerjee:2019pks,Baglio:2020oqu,Rao:2020hel}.

The loop-induced gluon-initiated contributions, calculated at LO in Refs.~\cite{Dicus:1988yh,Kniehl:1990iva}, are finite and enter at order $\alpha_s^2$, i.e. formally at NNLO considering the $pp\to VH$ process.
However, due to the dominance of the gluon PDFs at the LHC they are sizeable, contributing about 6\% of the total NNLO cross section, and the contribution can be twice as large in the boosted Higgs boson regime $p_T^H\gtrsim 150$\,GeV~\cite{Harlander:2013mla,Englert:2013vua}.
As only the LO results for  $gg\to ZH$ are known and implemented in current Monte Carlo generators, the large scale uncertainties pertaining to this channel constitute a significant part of the uncertainties in experimental analyses of the $VH$ process~\cite{Aaboud:2018zhk,Sirunyan:2018kst,Aad:2020eiv,Sirunyan:2018cpi}.
Furthermore, the gluon-initiated subprocess is very sensitive to modified Yukawa couplings and/or non--SM particles running in the loop,
For these reasons the NLO corrections to this process are very important.
However, the NLO corrections  contain two-loop integrals involving $m_t, m_H$ and $m_Z$, and such integrals are currently unknown analytically.
Therefore the QCD corrections  have been calculated in various approximations so far.
In Ref.~\cite{Altenkamp:2012sx} they have been calculated in the $m_t\to \infty$ limit, resulting in a K-factor used to reweight the full one-loop result.
In addition, top quark mass effects at NLO QCD have been considered in the framework of a $1/m_t$-expansion in Ref.~\cite{Hasselhuhn:2016rqt}, including Pad\'e approximants constructed from expansion terms up to $1/m_t^8$.
However, the $1/m_t$-expansion becomes invalid for invariant masses of the $HZ$ system larger than $2m_t$, which is the interesting region with regards to new physics searches.
Soft gluon resummation for the $gg\to ZH$ process has been calculated in Ref.~\cite{Harlander:2014wda}.


In Ref.~\cite{Harlander:2018yns}, a data-driven strategy to extract the gluon-initiated component
(or, more precisely, the non-Drell-Yan component)
for $ZH$ production has been suggested, 
based on the comparison of the $ZH$ to the $WH$ cross section and the corresponding invariant mass distributions of the $VH$ system.


Considering the full process $pp\to ZH$, 
inclusive NNLO  QCD corrections have been available for quite some time~\cite{Brein:2003wg} and are implemented 
in the program {\sc VH@NNLO}~\cite{Brein:2003wg,Brein:2011vx,Brein:2012ne,Harlander:2018yio}.
NLO electroweak corrections have been calculated in Refs.~\cite{Ciccolini:2003jy,Denner:2011id}, combined NLO QCD+EW corrections are also available~\cite{Denner:2014cla,Granata:2017iod,Obul:2018psx}.

Differential QCD corrections to the process $pp\to ZH$ have  been calculated up to NNLO, including $H\to b\bar{b}$ decays at NLO~\cite{Ferrera:2014lca,Campbell:2016jau} and at NNLO~\cite{Ferrera:2017zex}. 
In Ref.~\cite{Gauld:2019yng},  fully differential NNLO results for $VH$ observables including the decays of the vector boson into leptons and the Higgs boson into $b$-quarks with off-shell propagators of the vector- and Higgs bosons have been calculated.

The combination of fixed-order QCD computations with parton showers has been studied at NLO+PS in association with up to one jet~\cite{Luisoni:2013cuh,Goncalves:2015mfa,Hespel:2015zea,Granata:2017iod} and at NNLOPS~\cite{Astill:2018ivh,Bizon:2019tfo}.
In addition, the NNLO corrections have been combined with NNLL resummation in the 0-jettiness variable and matched
to a parton shower within the {\sc Geneva} Monte Carlo framework~\cite{Alioli:2019qzz}.

Threshold corrections up to N$^3$LL for the Drell-Yan-type part of the  inclusive cross section have been calculated in Ref.~\cite{Kumar:2014uwa}, 
soft-gluon resummation of both threshold logarithms and logarithms which are important at low transverse momentum of the $V H$ pair have been considered up to N$^3$LL in Ref.~\cite{Dawson:2012gs} and have been found to be very close to the fixed order NNLO result.
The process $b\bar{b}\to ZH$ in the five-flavour scheme, but with a non-vanishing
bottom-quark Yukawa coupling, has been calculated in the soft-virtual approximation at NNLO QCD in Ref.~\cite{Ahmed:2019udm}, the polarised $q\bar{q}\to ZH$ two-loop amplitudes have been calculated in Ref.~\cite{Ahmed:2020kme}.


In this paper, we calculate the two-loop virtual corrections to the process $gg\to ZH$, including massive top quarks in the loops.
We focus on the description of the calculation and display numerical results for the two-loop amplitudes.
A phenomenological study based on these results is postponed to a subsequent publication.
The structure of this work is as follows. In Section \ref{sec:calculation} we describe details of the calculation such as the integral families, our projection operators, the reduction to master integrals and the evaluation of the master integrals.
In Section \ref{sec:results} we show results for the finite part of the virtual amplitude, for some benchmark points as well as in terms of two-dimensional grids, before we conclude in Section \ref{sec:conclusions}.

\section{Details of the calculation}
\label{sec:calculation}

\subsection{Amplitude definition}

\begin{figure}
\centering
\includegraphics[width=\textwidth]{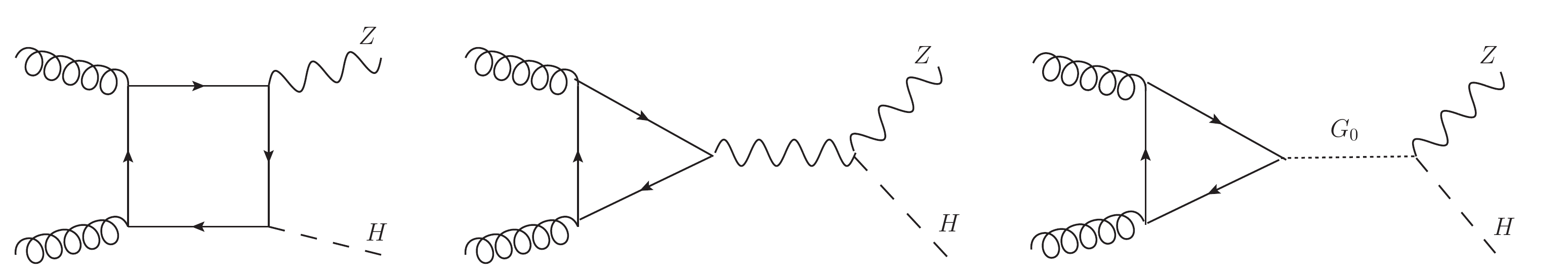}
\caption{Diagrams contributing to $gg\to ZH$ at leading order. Diagrams related by crossings are not shown.\label{fig:diagsLO}}
\end{figure}

We consider the process
\begin{align}\label{eq:process}
g(p_1) + g(p_2) \to Z(p_3) + H(p_4) ,
\end{align}
with $p_1^2 = p_2^2 = 0$, $p_3^2=\mz^2$ and $p_4^2=\mh^2$.  
The Mandelstam invariants are defined by
\begin{align} \label{eq:invariants}
s = \left(p_1 + p_2 \right)^2\;,\;
t = \left(p_2 - p_3 \right)^2\;,\;
u = \left(p_1 - p_3 \right)^2\;,
\end{align}
with $s+t+u=\mz^2+\mh^2$.
The diagrams contributing to the process (\ref{eq:process}) at leading order are shown in Fig.~\ref{fig:diagsLO}.
In our calculation we use the 't Hooft-Feynman gauge, therefore the diagram
involving the exchange of a would-be Goldstone boson $G_0$ needs to be taken
into account. We treat all quarks except the top-quark as massless, therefore only
top-quark loops contribute in those diagrams where the Higgs boson
couples directly to the fermion loop.
The effect of a finite bottom quark mass on the total LO cross section
is at the per mille level~\cite{Hasselhuhn:2016rqt}.

In the triangle diagrams, only the axial vector part of the $Z$-boson coupling contributes due to Furry's theorem.
In addition, the massless quark contributions cancel in each isospin
doublet, such that only the third generation of quarks give a non-zero 
contribution for this class of diagram.

The leading order amplitude can be expressed in terms of seven form
factors~\cite{Kniehl:1990iva} containing one-loop three- and
four-point functions. Some of the form factors can be related by
crossing $p_1\leftrightarrow p_2$ such that only four form factors
remain to be calculated. However we choose not to express our
amplitude in terms of these form factors, as will be explained
in Section \ref{sec:projection}.

In the $\mt\to\infty$ limit the amplitude simplifies considerably and is given by~\cite{Altenkamp:2012sx}
\begin{align}
M_0 =
-\frac{\als\alpha}{\sin^2\theta_{\mathrm{w}}\cos^2\theta_{\mathrm{w}}\,m_Z} \,\delta^{ab}
\eps(\varepsilon_1,\varepsilon_2,p_1,p_2) \,\frac{p_4\cdot\varepsilon_Z^*}{s}\;,
\label{eq:Born} 
\end{align}
where $\eps$ is the totally antisymmetric Levi-Civita symbol with $\eps(a,b,c,d)\equiv \eps^{\mu\nu\rho\sigma}a_\mu b_\nu c_\rho d_\sigma$ and $\varepsilon_1,\varepsilon_2$ are the polarisation vectors of the incoming gluons, carrying colour indices $a$ and $b$, while $\varepsilon_Z^*$ is the polarisation vector of the outgoing $Z$-boson.
The spin- and colour-averaged Born matrix element then can be written as
\begin{align}
\overline{|M_0|^2} = \frac{\als^2\alpha^2}{256\sin^4\theta_{\mathrm{w}}\cos^4\theta_{\mathrm{w}}\mz^4}\, \lambda(s,\mz^2,\mh^2)\;,
\label{eq:Bornsq} 
\end{align}
where $\lambda$ denotes the K{\"a}ll{\'e}n function, $\lambda(a,b,c)=a^2+b^2+c^2-2ab-2ac-2bc$.  

We calculate using conventional dimensional regularisation (CDR) with $D=4-2\epsilon$.
For the treatment of $\gamma_5$ within dimensional regularisation we use the 't~Hooft--Veltman scheme~\cite{tHooft:1972tcz,Breitenlohner:1977hr} in the variant of Refs.~\cite{Larin:1991tj,Larin:1993tq}, i.e. we use
\begin{align}
  \gamma_5&=\frac{i}{4!}\eps_{\mu\nu\rho\sigma}\gamma^\mu\gamma^\nu\gamma^\rho\gamma^\sigma\;,\nn\\
  \frac{1}{2}(\gamma_\mu\gamma_5-\gamma_5\gamma_\mu)&=\frac{i}{3!}\eps_{\mu\nu\rho\sigma}\gamma^\nu\gamma^\rho\gamma^\sigma
\end{align}
and add finite renormalisation terms to restore chiral symmetry in the massless quark limit, see Section \ref{sec:renorm}.
The contraction of two $\eps$-symbols leads to linear combinations of metric tensors which are treated as $D$-dimensional.

\subsubsection{Tensor structures and projection to a basis of linear polarisations}
\label{sec:projection}

We define the tensor amplitude $\amp_{\mu_1\mu_2\mu_3}$ by extracting the polarisation vectors from the amplitude of process (\ref{eq:process}),
\begin{align} \label{eq:tensoramplitudes}
\amp{} &= \varepsilon_{\lambda_1}^{\mu_1}(p_1)\,\varepsilon_{\lambda_2}^{\mu_2}(p_2)\,(\varepsilon_{\lambda_3}^{\mu_3}(p_3))^\star\,\amp_{\mu_1\mu_2\mu_3} \;,
\end{align} 
where the $\varepsilon_{\lambda_i}^{\mu_i}$ denote the polarisation vectors.
The Lorentz tensor structures appearing in the amplitude $\amp_{\mu_1\mu_2\mu_3}$  were discussed in Ref.~\cite{Kniehl:1990iva}.
However, we do not use form factors related to these Lorentz structures here, but rather use projections based on the momentum-basis representations of the linear polarisation vectors of external particles as suggested in Ref.~\cite{Chen:2019wyb}. 
We also use the fact that  only one axial current is involved in the QCD corrections to this amplitude, therefore all relevant Lorentz structures
contain only a single Levi-Civita symbol.
In addition, conditions such as transversality and Bose symmetry regarding the two external gluons further constrain the possible Lorentz structures. 

\begin{figure}[tbh!]
\centering
\includegraphics[width=11cm,height=5cm]{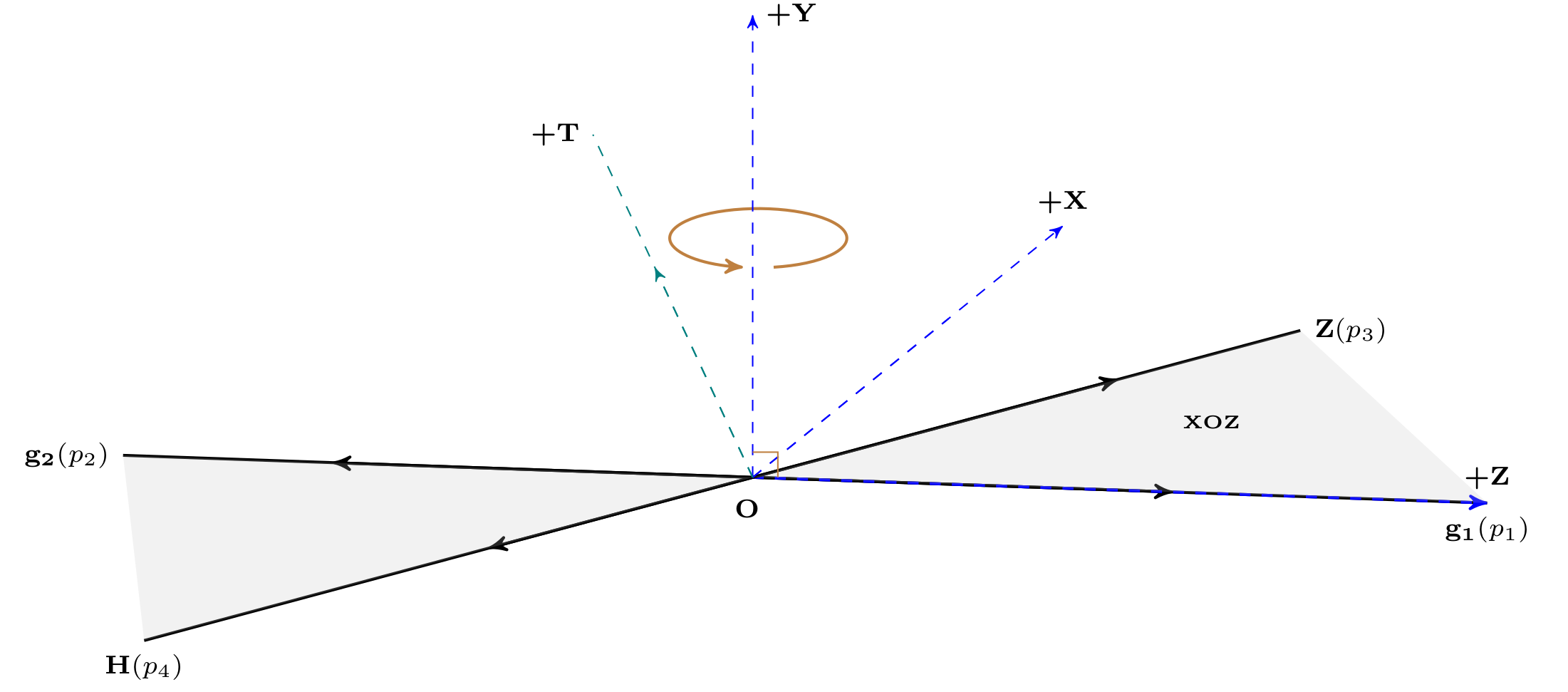}
\caption{The coordinate system in the center-of-mass frame of the two incoming gluons.}
\label{FIG:coordinatesystem}
\end{figure}

Following the procedure of Ref.~\cite{Chen:2019wyb}, we define the following normalised linear polarisation vectors, where the frame that has been used is shown in Fig.~\ref{FIG:coordinatesystem},
\begin{eqnarray} 
\label{eq:LPstatevectors}
\varepsilon^{\mu}_{x} &=& \mathcal{N}_x\, \left(-s_{23}  p^{\mu}_{1} - s_{13} p^{\mu}_2 + s_{12} p^{\mu}_{3}\right)\,,\nonumber\\ 
\varepsilon_{y}^{\mu} &=& \mathcal{N}_y\, \left(\epsilon_{\mu_1\,\mu_2\,\mu_3}^{\mu}\, p^{\mu_1}_1 \, p^{\mu_2}_2 \,p^{\mu_3}_3 \right), \nonumber\\
\varepsilon^{\mu}_{T} &=& \mathcal{N}_T\, \left( \left(-s_{23} (s_{13}
                          + s_{23})+2 \mz^2 s_{12}\right) p^{\mu}_{1}
                          + \left(s_{13} (s_{13} + s_{23})-2 \mz^2
                          s_{12} \right) p^{\mu}_2 \right.\nn\\
  &&\left. + s_{12} (-s_{13}+s_{23})\, p^{\mu}_{3} \,\right)\,,\nonumber\\ 
\varepsilon^{\mu}_{l} &=& \mathcal{N}_l\, \left(-2\mz^2 \left(p^{\mu}_1 + p^{\mu}_2\right) + \left(s_{13} + s_{23}\right) p^{\mu}_{3}\right)\,,
\end{eqnarray}
with $s_{12} \equiv 2 p_1 \cdot p_2 = s \,,\, s_{13} \equiv 2 p_1 \cdot p_3 = s+t-m_H^2\,,\, s_{23} \equiv 2 p_2 \cdot p_3 =\mz^2 - t$.
The normalisation factors $\mathcal{N}_i$ for $i \in \{x,y,T,l\}$,
associated with each of these space-like polarisation vectors in
Eq.~(\ref{eq:LPstatevectors}), can be determined from their (negative)
norm squares $\varepsilon^2_{i}$, i.e. $\mathcal{N}_i =
1/\sqrt{-\varepsilon^2_{i}}$, which we choose to include only at the very end of the calculation. 

The physical meaning of these vectors is apparent in the center-of-mass frame of $p_1$ and $p_2$, where the beam axis determined by $\left\{\vec{p}_1, \vec{p}_2 \right\}$ is taken as the $z$-axis and the plane determined by $\left\{\vec{p}_1, \vec{p}_3 \right\}$ defines the $x$-O-$z$ plane, as illustrated in Fig.~\ref{FIG:coordinatesystem}.
Then the polarisation vector $\varepsilon_{x}$ is orthogonal to the beam axis and lies within the $x$-O-$z$ plane, while $\varepsilon_{y}$ is perpendicular to this plane.
The vector $\varepsilon_{T}$ lies within the $x$-O-$z$ plane but points to a direction orthogonal to $\vec{p}_3$,
and $\varepsilon_{l}$, the longitudinal polarisation of the $Z$-boson,
points along the direction of $\vec{p}_3$ in the center-of-mass frame.

Our projectors are given by products of three linear polarisation vectors,
\begin{equation}
\label{eq:LPprojectorsraw}
\varepsilon^{\mu_1}_{i}\,\varepsilon^{\mu_2}_{j}\,\varepsilon^{\mu_3}_{k}\, \text{ with $i\,,j \in \{ x,y\}$ and $k \in \{T,y,l\}$},
\end{equation}
which are further re-written solely in terms of external momenta and the Levi-Civita symbol, all with $D$-dimensional Lorentz indices. 
However, only the six projectors with an odd number of Levi-Civita symbol, i.e.  an odd number of $\varepsilon_y$, are relevant for the process (\ref{eq:process}). This is because there are only three linearly independent external momenta in $\amp_{\mu_1\mu_2\mu_3}$ and all Lorentz structures appearing in the amplitude contain a Levi-Civita symbol.
Concretely, the six projectors we use for $\amp_{\mu_1\mu_2\mu_3}$ are given by
\begin{eqnarray}
\label{eq:LPprojectors}
\mathcal{P}^{\mu_1\mu_2\mu_3}_1 &=& \varepsilon^{\mu_1}_{x}\,\varepsilon^{\mu_2}_{x}\,\varepsilon^{\mu_3}_{y}\,, \quad \quad 
\mathcal{P}^{\mu_1\mu_2\mu_3}_2 = \varepsilon^{\mu_1}_{x}\,\varepsilon^{\mu_2}_{y}\,\varepsilon^{\mu_3}_{T}, \nonumber\\
\mathcal{P}^{\mu_1\mu_2\mu_3}_3 &=& \varepsilon^{\mu_1}_{x}\,\varepsilon^{\mu_2}_{y}\,\varepsilon^{\mu_3}_{l}\,, \quad \quad 
\mathcal{P}^{\mu_1\mu_2\mu_3}_4 = \varepsilon^{\mu_1}_{y}\,\varepsilon^{\mu_2}_{x}\,\varepsilon^{\mu_3}_{T}, \nonumber\\
\mathcal{P}^{\mu_1\mu_2\mu_3}_5 &=& \varepsilon^{\mu_1}_{y}\,\varepsilon^{\mu_2}_{x}\,\varepsilon^{\mu_3}_{l}\,, \quad \quad 
\mathcal{P}^{\mu_1\mu_2\mu_3}_6 = \varepsilon^{\mu_1}_{y}\,\varepsilon^{\mu_2}_{y}\,\varepsilon^{\mu_3}_{y}\,.
\end{eqnarray}
The projections made with $\mathcal{P}^{\mu_1\mu_2\mu_3}_4$ and $\mathcal{P}^{\mu_1\mu_2\mu_3}_5$ are related to those with $\mathcal{P}^{\mu_1\mu_2\mu_3}_2$ and $\mathcal{P}^{\mu_1\mu_2\mu_3}_3$, respectively, through crossing of $p_1$ and $p_2$.
Therefore we only need to compute four projections in total.
Note that all open Lorentz indices in Eq.~(\ref{eq:LPprojectors})  are understood to be $D$-dimensional, and pairs of Levi-Civita symbols in $\mathcal{P}^{\mu_1\mu_2\mu_3}_6$ need to be contracted \textit{before} being used for the projection of the amplitude (see Ref.~\cite{Chen:2019wyb} for a more detailed discussion).

From the point of view of a form factor decomposition, the set of linear-polarisation projectors in Eq.~(\ref{eq:LPprojectors}) represent precisely the Lorentz tensor decomposition basis in use.
This is a consequence of the orthogonality of the projectors (although the linear completeness is in general only ensured in the 4-dimensional limit~\cite{Chen:2019wyb,Ahmed:2019udm}).

This projection defines six quantities 
\begin{equation}
\label{eq:LPamps}
\amp_{n} \equiv \mathcal{P}^{\mu_1\mu_2\mu_3}_n \, \amp_{\mu_1\mu_2\mu_3}\quad (n=1, \ldots, 6).
\end{equation}
The physical interpretation of the quantities $\amp_{n}$ as linearly polarised amplitudes offers us a convenient short-cut
to  transform to a helicity basis defined w.r.t the same reference frame.
The usual helicity amplitudes for the process (\ref{eq:process}) can be constructed from the linear ones, using the relations
\begin{eqnarray}
\label{eq:LP2HL}
\varepsilon^{\mu_1}_{\pm}(p_1) &= \frac{1}{\sqrt{2}} \left( \varepsilon_x^{\mu_1} \pm i \varepsilon_y^{\mu_1} \right)\,,\nonumber\\
\varepsilon^{\mu_2}_{\pm}(p_2) &= \frac{1}{\sqrt{2}} \left( \varepsilon_x^{\mu_2} \mp i \varepsilon_y^{\mu_2} \right)\,,\nonumber\\
\varepsilon^{\mu_3}_{\pm}(p_3) &= \frac{1}{\sqrt{2}} \left( \varepsilon_T^{\mu_3} \pm i \varepsilon_y^{\mu_3} \right)\, 
\end{eqnarray}
and the longitudinal polarisation of the $Z$-boson is given by $\varepsilon^{\mu_3}_{l}$  in Eq.~(\ref{eq:LPstatevectors}).

\subsubsection{Kinematics}

The kinematic invariants in terms of the scattering angle $\theta$
between the beam axis and $\vec{p}_3$ in the centre-of-mass frame are
given by
\begin{align}
  t&=-\frac{1}{2}\left(s-\mz^2-\mh^2+\cos\theta\sqrt{\lambda(s,\mz^2,\mh^2)}   \right)\;,\nn\\
  u&=-\frac{1}{2}\left(s-\mz^2-\mh^2-\cos\theta\sqrt{\lambda(s,\mz^2,\mh^2)}   \right)\;,
\end{align}
where 
  $\lambda(s,\mz^2,\mh^2)= \left( s-(\mz+\mh)^2\right)\left( s-(\mz-\mh)^2\right)$ 
is the  K{\"a}ll{\'e}n function.
Defining $m_+=\mz+\mh, m_-=\mz-\mh$ and $\beta_{ZH}^2=\left( 1-\frac{m_+^2}{s}\right)\left( 1-\frac{m_-^2}{s}\right)$, we can also write
\begin{align}
  t&=\frac{1}{2}\left(\mz^2+\mh^2\right)-\frac{s}{2}\left( 1+\beta_{ZH}\cos\theta\right)\;,\nn\\
  u&=\frac{1}{2}\left(\mz^2+\mh^2\right)-\frac{s}{2}\left( 1-\beta_{ZH}\cos\theta\right)\;.
\end{align}
The limit $\beta_{ZH}\to 1$ corresponds to $s\gg m_+^2,m_-^2$. In the forward scattering region $\theta\to 0$, if in addition $\beta_{ZH}\to 1$ is fulfilled, we have $|t|\gg |u|$. Analogously, for $\theta\to \pi$ and $\beta_{ZH}\to 1$, the ratio $t/u$ is very small.

The virtual amplitude has a threshold at $s=4m_t^2$, therefore it is
also useful to define
\begin{align}
  \beta_t=\frac{s-4m_t^2}{s+4m_t^2-2 (\mz+\mh)^2}\;,
 \end{align} 
  such that $-1\leq \beta_t\leq 1$ with $\beta_t=-1$ for
  $s=s_{min}=(\mz+\mh)^2$, $\beta_t=0$ at the top quark pair
  production threshold $s=4m_t^2$ and $\beta_t\to 1$ for $s\to\infty$.

  For $2\to 2$ kinematics, the transverse momentum $p_T$ of the Higgs-
  or $Z$-boson (with $\vec{p}_T^{\,H}=-\vec{p}_T^{\,Z}$),
 fulfils $0\leq p_T\leq \sqrt{\lambda}/(2\sqrt{s})$ and
 $s\,p_T^2= t u -\mz^2\mh^2$. Therefore, in the high energy limit, $p_T^2 \to s\,(1-\cos^2\theta)$.
 
\vspace{3mm}
  

\subsection{Diagram Generation, reduction and calculation of the master integrals}

We generate the 1- and 2-loop diagrams using QGRAF~\cite{Nogueira:1991ex}. 
The projectors described in Section~\ref{sec:projection} are applied and the Feynman rules are inserted using FORM~\cite{Vermaseren:2000nd,Kuipers:2012rf,Ruijl:2017dtg}.
For the reduction  we have defined eight integral families ${\cal F}_i$,
five planar $(i=1,\ldots,5)$ and three non-planar $(i=6,\ldots,8)$ families.
We also use five additional families obtained from the original families by exchanging $p_1$ and $p_2$.
Each family contains nine propagators
which allows all irreducible scalar products in the numerator to be expressed in terms of inverse propagators.
Concretely, the occurring integrals have the form
\begin{equation}
I = (\mu_0^2)^{2\eps}\int \frac{\mbox{d}^D k_1}{i\pi^{\frac{D}{2}}} \int \frac{\mbox{d}^D k_2}{i\pi^{\frac{D}{2}}}  \,
N_{j_1}^{r_1} \ldots N_{j_t}^{r_t}
    N_{j_{t+1}}^{-s_1} \ldots N_{j_n}^{-s_{n-t}}\;,
\label{eq:Irs}
\end{equation}
where the $N_j$ denote genuine propagators of the form $1/(k^2-m^2)$ with exponents $r_i \geq 1$ and $s_i \geq 0$.

 The families are listed in Tables~\ref{tab:families_planar} and~\ref{tab:families_nonplanar}.
 Integral families which differ from the listed ones by exchanging
 $p_1$ and $p_2$ are not shown. The integrals appearing in the amplitude are reduced to a minimal set of master integrals as described in the following.

\begin{table}
\centering
\begin{tabular}{|l|l|l|}
\hline
${\cal F}_1$ & ${\cal F}_2$ & ${\cal F}_3$ \\
\hline
$k_1^2-m_t^2$ & $k_1^2-m_t^2$ & $k_1^2$ \\
$k_2^2-m_t^2$ & $k_2^2-m_t^2$ & $(k_1-k_2)^2-m_t^2$ \\
$(k_1-k_2)^2$ & $(k_1-k_2)^2$ & $(k_1+p_1)^2$ \\
$(k_1+p_1)^2-m_t^2$ & $(k_1+p_1)^2-m_t^2$ & $(k_2+p_1)^2-m_t^2$ \\
$(k_2+p_1)^2-m_t^2$ & $(k_2+p_1)^2-m_t^2$ & $(k_1-p_2)^2$ \\
$(k_1-p_2)^2-m_t^2$ & $(k_1-p_3)^2-m_t^2$ & $(k_2-p_2)^2-m_t^2$ \\
$(k_2-p_2)^2-m_t^2$ & $(k_2-p_3)^2-m_t^2$ & $(k_2-p_2-p_3)^2-m_t^2$ \\
$(k_1-p_2-p_3)^2-m_t^2$ & $(k_1-p_2-p_3)^2-m_t^2$ & $(k_1+p_1+p_3)^2$ \\
$(k_2-p_2-p_3)^2-m_t^2$ & $(k_2-p_2-p_3)^2-m_t^2$  & $(k_2+p_1-p_2)^2$ \\
\hline
\end{tabular}
\begin{tabular}{|l|l|}
${\cal F}_4$ & ${\cal F}_5$ \\
\hline
$k_1^2-m_t^2$ & $k_1^2$ \\
$k_2^2$ &  $k_2^2-m_t^2$ \\
$(k_1-k_2)^2-m_t^2$ & $(k_1-k_2)^2-m_t^2$ \\
$(k_1+p_1)^2-m_t^2$ &  $(k_1+p_1)^2$ \\
$(k_2+p_1)^2$ &  $(k_2+p_1)^2-m_t^2$ \\
$(k_1-p_2)^2-m_t^2$ & $(k_1-p_3)^2$ \\
$(k_2-p_2)^2$ & $(k_2-p_3)^2-m_t^2$ \\
$(k_1-p_2-p_3)^2-m_t^2$ & $(k_1-p_2-p_3)^2$ \\
$(k_2-p_2-p_3)^2$ & $(k_2-p_2-p_3)^2-m_t^2$ \\
\hline
\end{tabular}
\caption{Planar integral families used for the reduction.\label{tab:families_planar}}
\end{table}

\begin{table}
\centering
\begin{tabular}{|l|l|l|}
\hline
  ${\cal F}_6$ & ${\cal F}_7$ &${\cal F}_8$ \\
\hline
$k_1^2$ & $k_1^2-m_t^2$ &$k_1^2-m_t^2$ \\
$k_2^2$ &  $k_2^2-m_t^2$ &$k_2^2-m_t^2$ \\
$(k_1+p_1)^2$ &  $(k_1-k_2)^2$ &$(k_1-k_2)^2$\\
$(k_1-p_2)^2$ & $(k_1+p_3)^2-m_t^2$ &$(k_1-p_1-p_2-p_3)^2-m_t^2$\\
$(k_2+p_1)^2-m_t^2$ &  $(k_2+p_3)^2-m_t^2$  &$(k_2-p_1-p_2-p_3)^2-m_t^2$\\
$(k_1-k_2)^2-m_t^2$ & $(k_1-p_2)^2-m_t^2$&$(k_1-p_2)^2-m_t^2$\\
$(k_2-p_2-p_3)^2-m_t^2$ & $(k_2-p_2)^2-m_t^2$ &$(k_2-p_1)^2-m_t^2$\\
$(k_1-k_2+p_3)^2-m_t^2$ &$(k_1-k_2+p_1)^2$ &$(k_1-k_2+p_1)^2$ \\
$(k_1-k_2+p_2+p_3)^2$ & $(k_2-p_1-p_2)^2-m_t^2$ &$(k_2-p_1-p_2)^2-m_t^2$\\
\hline
\end{tabular}
\caption{Non-planar integral families used for the reduction.\label{tab:families_nonplanar}}
\end{table}

\subsubsection{Choice of master integrals and reduction procedure}

The numerical evaluation of finite integrals is typically much simpler than that of their divergent counterparts, we therefore follow the strategy of Ref.~\cite{vonManteuffel:2014qoa} to obtain a quasi-finite basis of master integrals. This choice of master integrals is not unique and the size of the coefficients of the integrals after reduction depends on this choice. In particular, it is known that it can be advantageous to choose a basis where the denominators in the reduction tables factorizes into factors containing only the space-time dimension~$D$ and factors depending on the kinematic invariants only. While algorithms to find such a basis have been presented in Refs.~\cite{Smirnov:2020quc,Usovitsch:2020jrk}, we obtained such a basis following a different approach, namely, by iterating over different combinations of master integral candidates and analysing the resulting reduction tables, restricted to a small subset of the full IBP system and neglecting subsectors. This allows us to define additional criteria  for the selection of the preferred masters, such as the size of the appearing denominator factors or the order in $\epsilon=(4-D)/2$ at which an integral starts contributing to the amplitude.

With our choice of master integrals, the $D$-dependence of the denominator factors of the reduction rules factors for all integrals, except for some one-particle reducible integrals. Furthermore, all seven-propagator integrals only start contributing to the amplitude at order $\epsilon^{-1}$ and the size of the amplitude coefficients reduces by about a factor of~5 compared to a default choice of finite masters with minimal propagator powers.
In particular the size of the coefficient of the two-propagator (double-tadpole) integral reduced from about 150\,MB to less than 5\,MB.

In order to perform the reduction to our chosen set of master integrals, we utilize the \texttt{Kira} package~\cite{Maierhoefer:2017hyi,Klappert:2020nbg} in combination with the rational function interpolation library \texttt{FireFly}~\cite{Klappert:2019emp,Klappert:2020aqs}. A crucial benefit of finite-field interpolation techniques is that one circumvents large intermediate expressions even during the calculation of IBP tables, which leads to a significant runtime and memory reduction in general. As a consequence, we interpolate the required reduction tables that relate the integrals of the amplitudes that contribute up to $r=7$ and $s=4$ to the desired set of master integrals that requires IBP relations up to $r=9$ and $s=2$ in $D$ dimensions.
To simplify the reduction of the integrals occurring in the amplitude, we scale the $Z$ and $H$ mass w.r.t. $\mt$ and approximate $\mz^2/\mt^2= 23/83$ and $\mh^2/\mt^2=12/23$, corresponding to $\mt=173.21$\,GeV, $\mz=91.18$\,GeV, and $\mh=125.1$\,GeV. Our value for $m_W$, which enters in electroweak couplings, amounts to $m_W=80.379\,$GeV. 
Note that, although we retain the dependence on $\mt$, the ratio of the $Z$-boson mass and Higgs boson mass to the top quark mass is fixed in our calculation. By setting $\mt=1$ we remove an additional scale that can be restored by dimensional analysis, which leaves us with a $3$--parameter problem.

As we use quasi-finite master integrals in six dimensions, the set of dimensional recurrence relations (DRRs), which connects integrals in six and four dimensions, has to be calculated as well. Hence, we utilize \texttt{LiteRed}~\cite{Lee:2013mka} and \texttt{Reduze}~\cite{vonManteuffel:2012np} in order to obtain these relations. The DRRs are subsequently related to our set of master integrals using \texttt{Kira} with \texttt{FireFly} in the same setup as described above. We note that the calculation of the relations between the DRRs and our basis of master integrals was the most demanding step in the whole calculation. It took about four days of wall-clock time running on a machine with two Intel\textsuperscript{\textregistered} Xeon\textsuperscript{\textregistered} Silver 4116 and hyper-threading. The required memory never exceeded about 100\,GB of memory.

Afterwards the DRRs and the reduction tables obtained by reducing the integrals of the amplitudes are combined to a custom system of equations that fills roughly 9\,GB of disk space. This system is again solved by employing \texttt{Kira} with \texttt{FireFly} in order to interpolate the final set of replacement rules. Finally, the latter set of replacements is inserted into the amplitudes with the help of the \texttt{ff\_insert} executable of \texttt{FireFly}.

It is worth mentioning that our calculation, which was split into
several substeps, could have been performed in a single run by using
the \texttt{Kira} option \texttt{iterative\_reduction: masterwise} with a custom system of equations that also includes the amplitudes. However, as one needs to hold the whole system of all integral families in memory at once in this case, we observed faster runtime and lower memory consumption by splitting the calculation as described above and running all steps on different machines in parallel. Additionally, splitting the reduction into several steps is convenient when studying different bases of master integrals and their impact of the total file size of the resulting amplitudes.

\subsubsection{Evaluation of the two-loop amplitude}

To evaluate the master integrals, we first apply sector decomposition as implemented in the program \pysecdec~\cite{Borowka:2017idc,Borowka:2018goh}.
For integrals which diverge in the limit of 4 space-time dimensions ($\epsilon \rightarrow 0$), sector decomposition resolves singularities in the regulator $\epsilon$ leaving only finite integrals over the Feynman parameters which can then be integrated numerically. In the physical region, singularities can appear on the real axis of the Feynman parameters and a causal prescription to avoid the singularities is required. We deform the integration contour of the Feynman parameters into the complex plane as described in Refs.~\cite{Soper:1999xk,Binoth:2005ff,Nagy:2006xy,Borowka:2012yc}.

In the present calculation we have selected a basis of quasi-finite integrals. This means that poles in $\epsilon$ can appear only in the prefactor of our integrals after Feynman parametrisation and, thus, sector decomposition is not required to resolve singularities in the regulator. Nevertheless, we observe that applying a sector decomposition greatly simplifies the numerical evaluation of the finite integrals. This observation can be partly understood by noting that sector decomposition also removes integrable singularities appearing at the boundaries where one or more Feynman parameters vanish. We therefore process all integrals with \pysecdec\ before numerical integration.

The numerical integration itself is performed using the quasi-Monte Carlo (QMC) algorithm~\cite{Li:2015foa,Borowka:2018goh}. For all integrals we apply a Korobov periodising transform with weight 3. We observe that the use of the rank-1 shifted lattice rules greatly reduces the number of samples required to obtain the integrals to sufficient precision for the computation of our amplitude, as compared to straightforward Monte Carlo sampling.

In line with our previous work on the processes $pp \rightarrow HH$~\cite{Borowka:2016ehy,Borowka:2016ypz}, $pp \rightarrow HJ$~\cite{Jones:2018hbb} and $gg \rightarrow \gamma \gamma$~\cite{Chen:2019fla}, we extend the \pysecdec\ program such that it can produce a code capable of evaluating the entire amplitude, rather than computing each integral separately. The advantage of this structure is that the number of sampling points used for each sector of each integral can be dynamically set according to its contribution to the total uncertainty on the amplitude. We utilise a variant of the procedure described in Ref.~\cite{Borowka:2016ypz} to minimise the total time taken to obtain a given relative accuracy on the amplitude.

\subsubsection{Renormalisation}
\label{sec:renorm}

We may expand each of the form factors, $\mathcal{A}_{i=1,\ldots,n}$, in the bare strong coupling $a_0 = \alpha_0/(4\pi)$ according to
\begin{equation}
\mathcal{A}_i = a_0 \mathcal{A}^{(0)}_i +a_0^2 \mathcal{A}^{(1)}_i + \mathcal{O}(a_0).
\label{eq:Aexpand}
\end{equation}
We renormalise the strong coupling in the $\overline{\mathrm{MS}}$ scheme with the heavy quark loop in the gluon self-energy subtracted at zero momentum. The heavy quark mass is renormalised in the $\mathrm{OS}$ scheme. The UV renormalised amplitude is obtained via the replacement
\begin{align}
\mathcal{A}^\mathrm{UV}_i &= Z_5 Z_A^{n_g/2} \mathcal{A}_i(a_0 \rightarrow a_s\,S_\epsilon^{-1} (\mu_R^2/\mu_0^2)^\epsilon\,Z_\alpha, m_0 \rightarrow m\,Z_m, y_{t0} \rightarrow y_{t}\,Z_m ) \nonumber \\
 =& 
 a_s\,  S_\epsilon^{-1} \left(\frac{\mu_R^2}{\mu_0^2}\right)^\epsilon \mathcal{A}_i^{(0)} 
 + a_s^2\, S_\epsilon^{-1} \left(\frac{\mu_R^2}{\mu_0^2}\right)^\epsilon (\frac{n_g}{2} \delta Z_A + \delta Z_\alpha + \delta Z_5) \mathcal{A}_i^{(0)} \nonumber\\ &  
 + a_s^2\, S_\epsilon^{-1} \left(\frac{\mu_R^2}{\mu_0^2}\right)^\epsilon \delta Z_m \mathcal{A}_i^{\mathrm{mct},(0)}  
 + a_s^2\, S_\epsilon^{-2} \left(\frac{\mu_R^2}{\mu_0^2}\right)^{2\epsilon} \mathcal{A}^{(1)} + \mathcal{O}(a_s^3)
\end{align}
Here, $n_g=2$ is the number of external gluon legs, $a_s=\alpha_s(\mu_R^2)/(4\pi)$ where $\alpha_s(\mu_R^2)$ is the renormalised coupling, $S_\epsilon=(4\pi)^\epsilon\,e^{-\epsilon \gamma_E}$, and the renormalisation constants are expanded according to
$Z_j= 1 + a_s \delta Z_j + \mathcal{O}(a_s^2)\; (j=\alpha,A,m,5)$,  with
\begin{align}
\delta Z_\alpha &= -\frac{1}{\epsilon} \beta_0 + \delta Z_\alpha^\mathrm{hq},\qquad \beta_0 = \frac{11}{3} C_A - \frac{4}{3} T_R\, n_F, \\
\delta Z_A &= - \delta Z_\alpha^\mathrm{hq} = \left(\frac{\mu_R^2}{m^2}\right)^\epsilon \left( -\frac{4}{3\epsilon} T_R \right), \\
\delta Z_m &= \left(\frac{\mu_R^2}{m^2}\right)^\epsilon C_F \left( -\frac{3}{\epsilon} - 4 \right),\\
\delta Z_5 &= \begin{cases}
\delta Z_{5,ns} &= -4 C_F, \quad \textrm{diagrams involving a Z-top vertex}\\
\delta Z_{5,\rho} &= -8 C_F, \quad \textrm{diagrams involving a Goldstone-top vertex}.
\end{cases}
\end{align}
We handle the $\gamma_5$ matrices which appear in this calculation using the Larin scheme~\cite{Larin:1993tq}. According to this scheme an additional finite renormalisation is required, it is denoted in the equations above by $\delta Z_5$.

At the level of individual form factors, the procedure outlined corresponds to the following relations
\begin{align}
\mathcal{A}^\mathrm{UV}_i  &= a_s \mathcal{A}_i^{(0),\mathrm{UV}}  + a_s^2 \mathcal{A}_i^{(1),\mathrm{UV}} + \mathcal{O}(a_s^3), \nonumber \\
\mathcal{A}_i^{(0),\mathrm{UV}}  &= S_\epsilon^{-1} \left(\frac{\mu_R^2}{\mu_0^2} \right)^\epsilon \mathcal{A}_i^{(0)}, \nonumber \\
\mathcal{A}_i^{(1),\mathrm{UV}}  &= S_\epsilon^{-2} \left(\frac{\mu_R^2}{\mu_0^2} \right)^{2\epsilon} \mathcal{A}_i^{(1)} + \left(-\, \frac{\beta_0}{\epsilon}+ \delta Z_5\right)\,\mathcal{A}_i^{(0),\mathrm{UV}} + S_\epsilon^{-1} \left(\frac{\mu_R^2}{\mu_0^2}\right)^\epsilon \delta Z_m\,\mathcal{A}_i^{\mathrm{mct},(0)}.
\end{align}

\subsubsection{Definition of the finite part of the virtual two-loop amplitude}

In order to obtain IR finite amplitudes we use the subtraction scheme described in Ref.~\cite{Catani:2013tia}
\begin{align}
\mathcal{A}_i^{(0),\mathrm{fin}} &= \mathcal{A}_i^{(0),\mathrm{UV}}, \\
\mathcal{A}_i^{(1),\mathrm{fin}} &=  \mathcal{A}_i^{(1),\mathrm{UV}} - I_1  \mathcal{A}_i^{(0),\mathrm{UV}},
\end{align}
with
\begin{align}
I_1 &= I_1^\mathrm{soft} + I_1^\mathrm{coll}, \\
I_1^\mathrm{soft} &= -\frac{e^{\epsilon \gamma_E}}{\Gamma(1-\epsilon)} \left(\frac{\mu_R^2}{s}\right)^\epsilon \left( \frac{1}{\epsilon^2} + \frac{i \pi}{\epsilon} \right) 2 C_A, \\
I_1^\mathrm{coll} &= - \frac{\beta_0}{\epsilon} \left(\frac{\mu_R^2}{s}\right)^\epsilon.
\end{align}
We present results at renormalisation scale $\mu_R^2 = s$ and after changing to the helicity basis, we define
\begin{align}
  \mathcal B&= \frac{8\, T_R^2}{(2\cdot 8)^2} \sum_\mathrm{\{\lambda_i\}}\mathcal{A}_{\{\lambda_i\}}^{(0),\mathrm{fin}}\mathcal{A}_{\{\lambda_i\}}^{\star(0),\mathrm{fin}},\\
  \mathcal V&= \frac{8\, T_R^2}{(2\cdot 8)^2}\sum_\mathrm{{\{\lambda_i\}}}\left( \mathcal{A}_{{\{\lambda_i\}}}^{(0),\mathrm{fin}}\mathcal{A}_{{\{\lambda_i\}}}^{\star(1),\mathrm{fin}}
                +\mathcal{A}_{{\{\lambda_i\}}}^{(1),\mathrm{fin}}\mathcal{A}_{{\{\lambda_i\}}}^{\star(0),\mathrm{fin}}\right)
\end{align}
for the square of the amplitude. 
We set the overall factor of $e^2 = 4\pi\alpha$ to one in our amplitude.

\section{Results}
\label{sec:results}

\subsection{Checks of the calculation}

\begin{figure}
\centering
\includegraphics[width=0.8\textwidth]{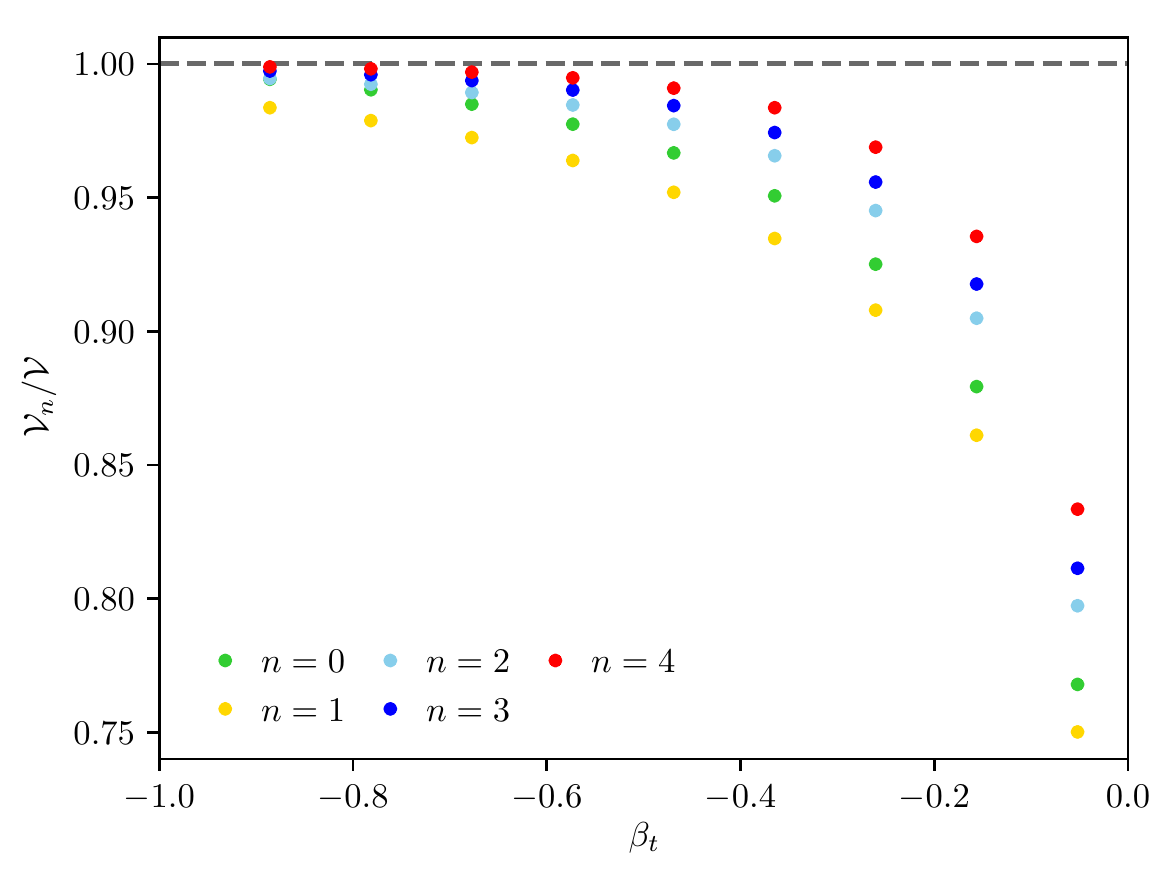}
\caption{Comparison of exact virtual contributions $\mathcal{V}$ with the expansion in the large top quark mass limit $\mathcal{V}_n$ from~\cite{Hasselhuhn:2016rqt} for fixed scattering angle $\mathrm{cos}(\theta) \approx 0.052$.
The range of the horizontal axis $-1\leq\beta_t\leq 0$ corresponds to the energy range $\left(\mz+\mh\right)^2 \leq s \leq 4\mt^2$ between production threshold and top quark pair threshold.}
\label{fig:expcomparison}
\end{figure}

We have verified that our results have the expected universal pole structure and crossing symmetries.
We also compared our exact result with the expansion in the large top quark mass limit presented in~\cite{Hasselhuhn:2016rqt}.
The virtual contributions $\widetilde{\mathcal{V}}_n$ are expanded up to order $1/\mt^{2n}$ and reweighted with the exact Born term $\mathcal{B}$.
The one-particle-reducible double triangle contributions $\mathcal{V}^{\mathrm{red}}$ are included with full top quark mass dependence.
\begin{equation}
\mathcal{V}_n =  \frac{\mathcal{B}}{\mathcal{B}_n}\widetilde{\mathcal{V}}_n + \mathcal{V}^{\mathrm{red}}
\end{equation}
In Fig.~\ref{fig:expcomparison} the ratio of expanded to full virtual contribution $\mathcal{V}_n/\mathcal{V}$ is shown for expansion orders $0\leq n \leq 4$ for a fixed scattering angle $\mathrm{cos}(\theta) \approx 0.052$.
For energies close to the production threshold at $\beta_t = -1$
($s=\left(\mz+\mh\right)^2$) the expanded result approximates the
exact calculation well with a ratio $\mathcal{V}_4/\mathcal{V} \approx
0.9989$, while the agreement worsens closer to the top quark pair
threshold at $\beta_t=0$ ($s=4\mt^2$), where the large $\mt$-expansion is expected
to break down.
 
In addition, we have compared our results to a recent calculation in the high energy limit~\cite{Davies:2020drs} and found agreement to the extent expected from previous comparisons for the process $pp\to HH$ performed in Ref.~\cite{Davies:2019dfy}.

\subsection{Numerical results for the two-loop amplitudes}

\begin{figure}
\centering
\includegraphics[width=0.49\textwidth]{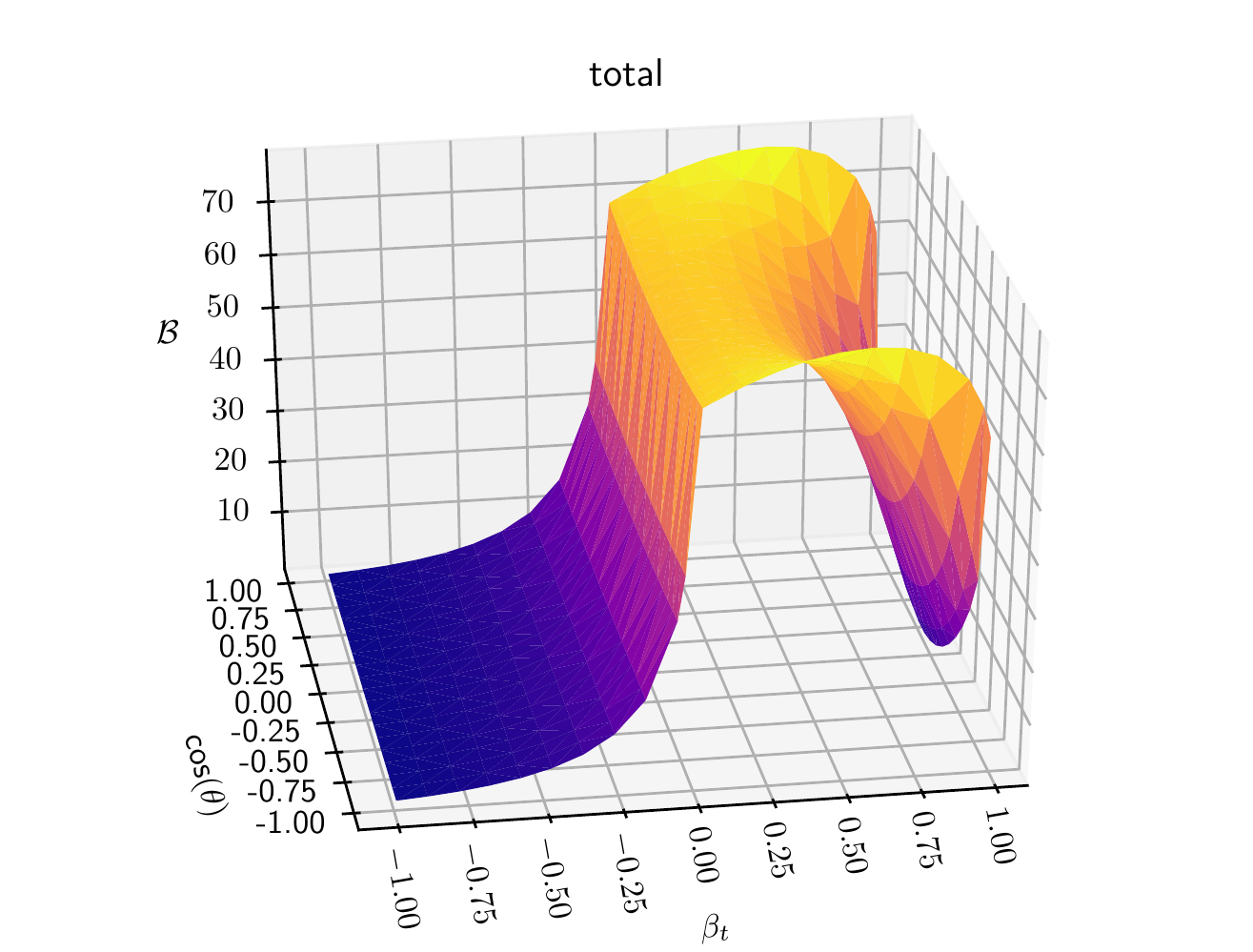}
\includegraphics[width=0.49\textwidth]{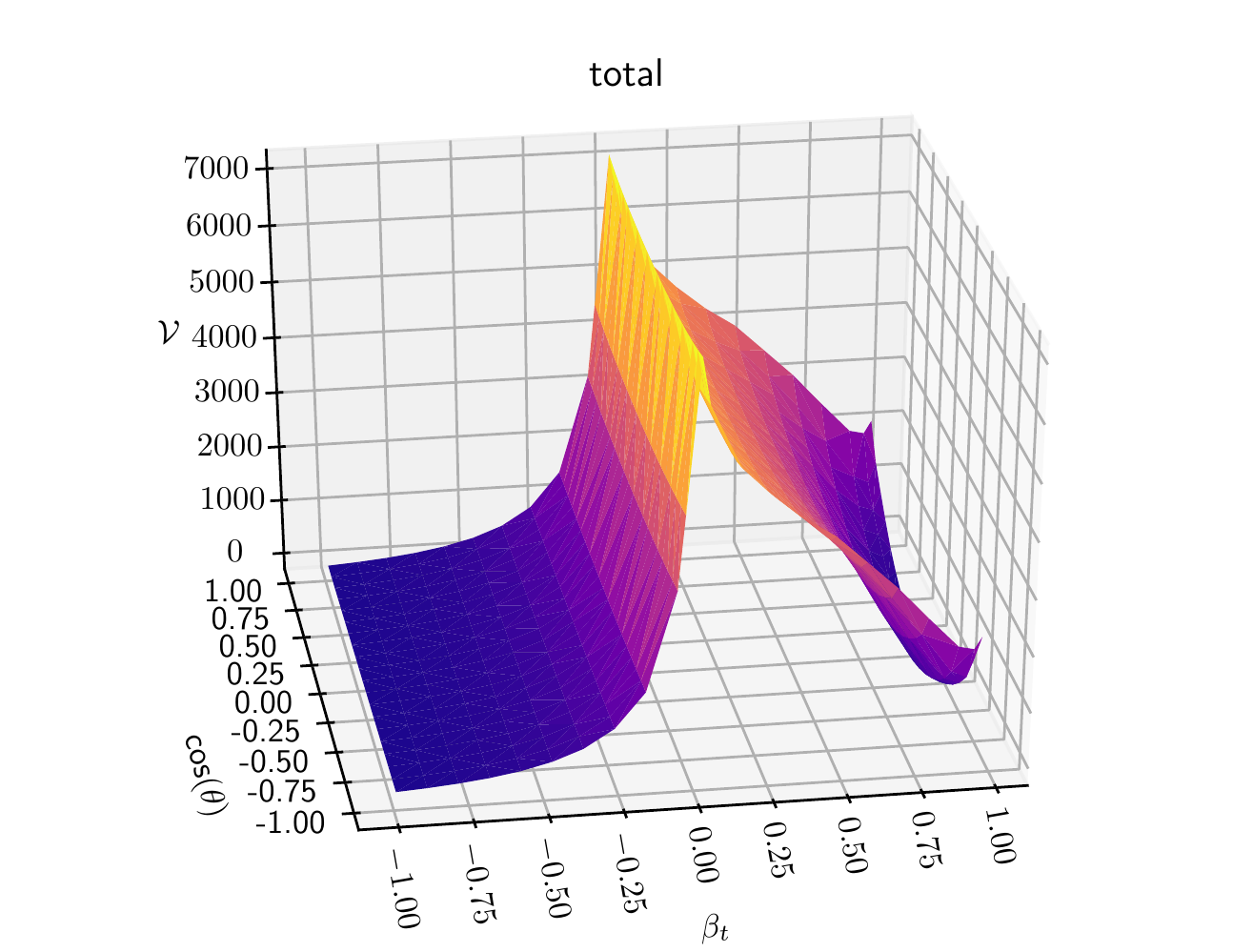}
  \caption{Dependence of the leading order~(left) and virtual contribution~(right) on the parameters $\beta_t$ and $\cos\theta$, summing over all polarisation.  }
\label{fig:unpoarised}
\end{figure}

\begin{figure}
\centering
\includegraphics[width=0.49\textwidth]{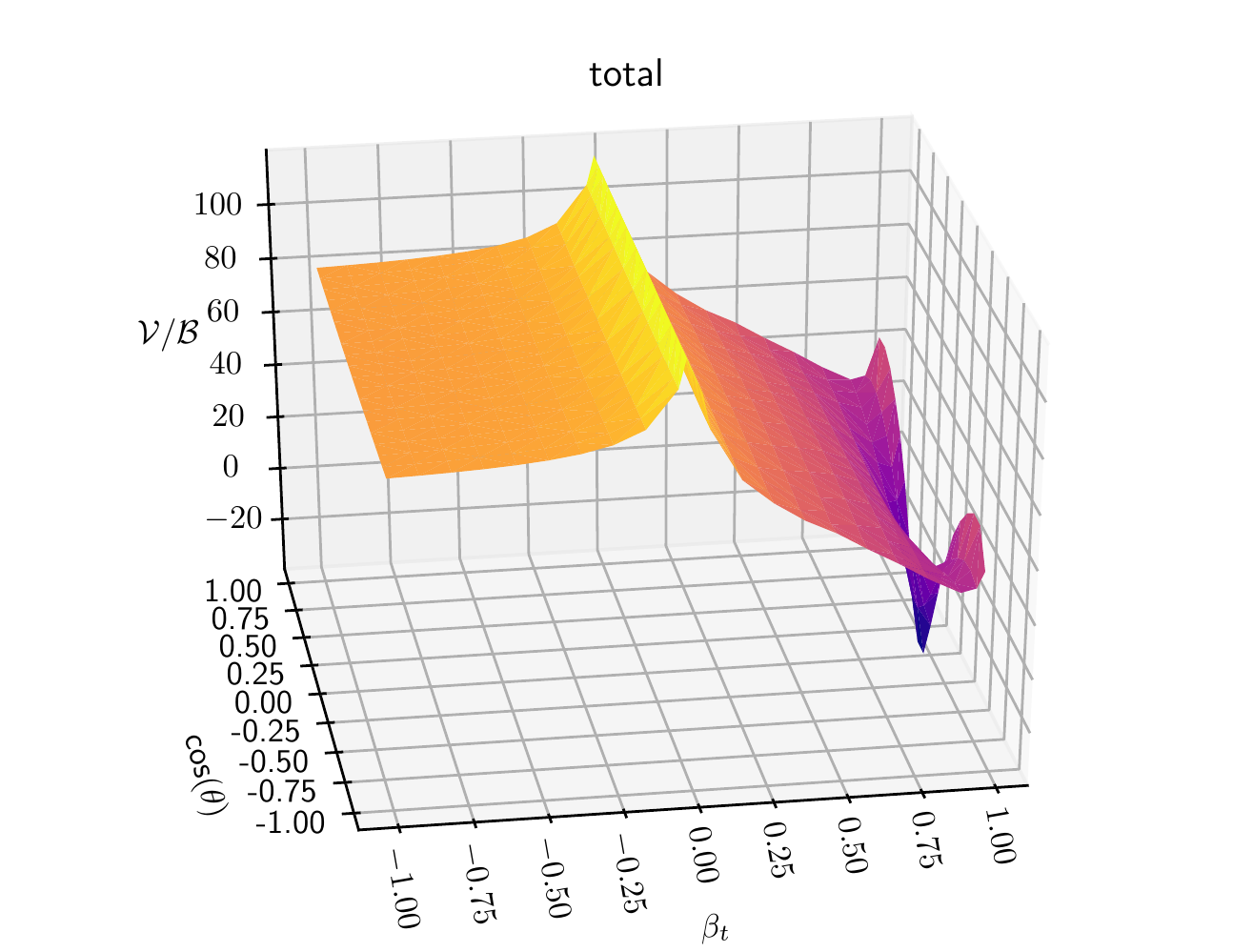}
  \caption{Dependence of the ratio $\mathcal V/\mathcal B$ on the parameters $\beta_t$ and $\cos\theta$, summing over all polarisations.  }
\label{fig:kfac}
\end{figure}

For the presentation of our results, we have evaluated a total of 460 phase-space points at 2-loop. We request per mille precision for each of the linearly polarised amplitudes, this is obtained for most phase-space points with between 45 minutes and 24 hours of run time using 2 x Nvidia Tesla V100 Graphics processing units (GPUs).
In Fig.~\ref{fig:unpoarised}, we show results for the unpolarised
modulus of the Born amplitude, as well as for the finite part of the virtual two-loop
amplitude interfered with the Born amplitude. 
We see that the unpolarised Born result shows a rather flat dependence
on the scattering angle around and below the top quark pair threshold,
due to the dominating s-wave contribution in this region, and starts
to curve as energy further increases because partial waves with higher
angular momentum play an increasingly important role.
We also observe that in the two-loop case, the top quark pair production threshold region is much more
peaked than at leading order, due to $\log\beta_t$-terms appearing for
the first time at two-loop order. In Fig.~\ref{fig:kfac}, the ratio of the two-loop amplitude to the Born-amplitude is shown separately.

In the following we will show results for five helicity amplitudes in the $\beta_t$--$\cos\theta$--plane.
The remaining amplitudes can be obtained from overall helicity flips as well as crossing between the two initial gluons.
Bose-symmetry and the behaviour under parity transformations imply
\begin{align}
  {\mathcal A}_{\lambda_1,\lambda_2,\lambda_z}(t,u)&= (-1)^{\lambda_z}{\mathcal A}_{\lambda_2,\lambda_1,\lambda_z}(u,t)\;,\label{eq:bose}\\
  {\mathcal A}_{\lambda_1,\lambda_2,\lambda_z}(t,u)&= -{\mathcal A}_{-\lambda_1,-\lambda_2,-\lambda_z}(t,u)\;.\nn
\end{align}
To understand the qualitative behaviour, we can expand the amplitude in terms of Wigner $d$-functions,
\begin{align}
 {\mathcal A}_{\{\lambda_i\}} (\theta) = \sum_{J=0}^{\infty}
 (2J+1){\mathcal A}^{J}_{\{\lambda_i\}} d^{J}_{s_i, s_f}(\theta)\;,
 \label{eq:partialwaves}
\end{align}
where $s_i$ and $s_f$ denote the initial and final state total spins,
respectively, and $J$ denotes the total angular momentum of the system.
For $gg \rightarrow ZH$,  the initial state has total spin $s_i=0$ for equal helicities of the initial state gluons,
$(\lambda_1,\lambda_2)=(+,+)$ or $(-,-)$, while for the case $(\lambda_1,\lambda_2)=(+,-)$ or $(-,+)$,  the initial state has total spin $s_i=2$.
Therefore the amplitude ${\cal A}_{++0}$ is dominated by the partial wave $d^0_{00}(\theta)$ and provides the largest contribution to the total squared amplitude.
In particular in the low energy region, where the $ZH$ system has relatively small kinetic energy, the s-wave contribution should dominate, reflected in the homogeneity of ${\cal A}_{++0}$ in $\cos\theta$.
As the center-of-mass energy increases,  the contributions from partial waves with higher angular momenta also start to play a role, leading to a non-flat behaviour in $\cos\theta$.
Note that eq.~(\ref{eq:bose}) implies that ${\cal A}_{++0}$ is
symmetric under exchange of $t\leftrightarrow u$ and therefore is
symmetric in $\cos\theta$. As  ${\cal A}_{++0}$ is composed of partial
waves  $d^J_{00}(\theta)$, which are even in $\cos\theta$ for even
$J$, no partial wave components with odd $J$ can contribute to ${\cal A}_{++0}$.

\begin{figure}
\centering
\includegraphics[width=0.49\textwidth]{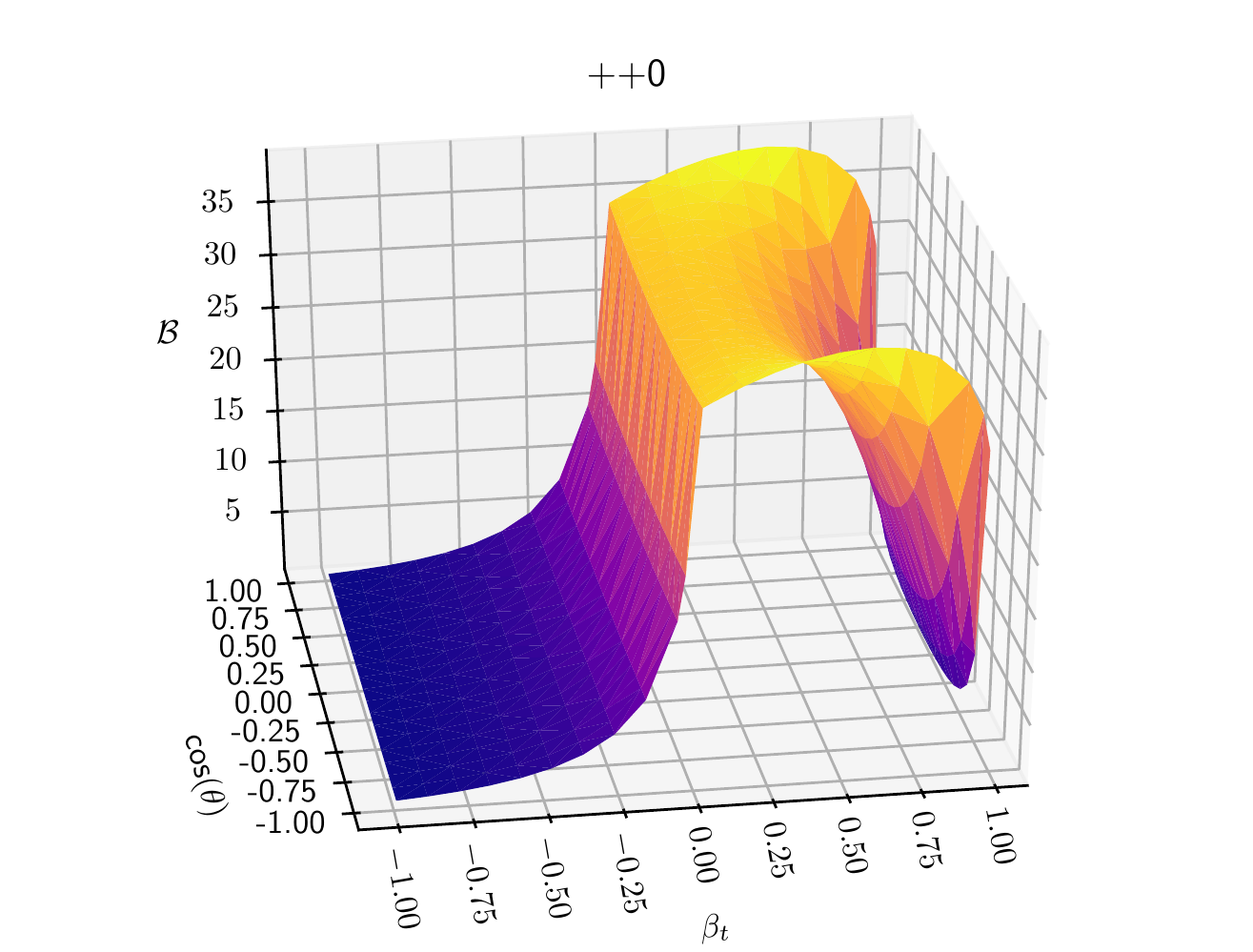}
\includegraphics[width=0.49\textwidth]{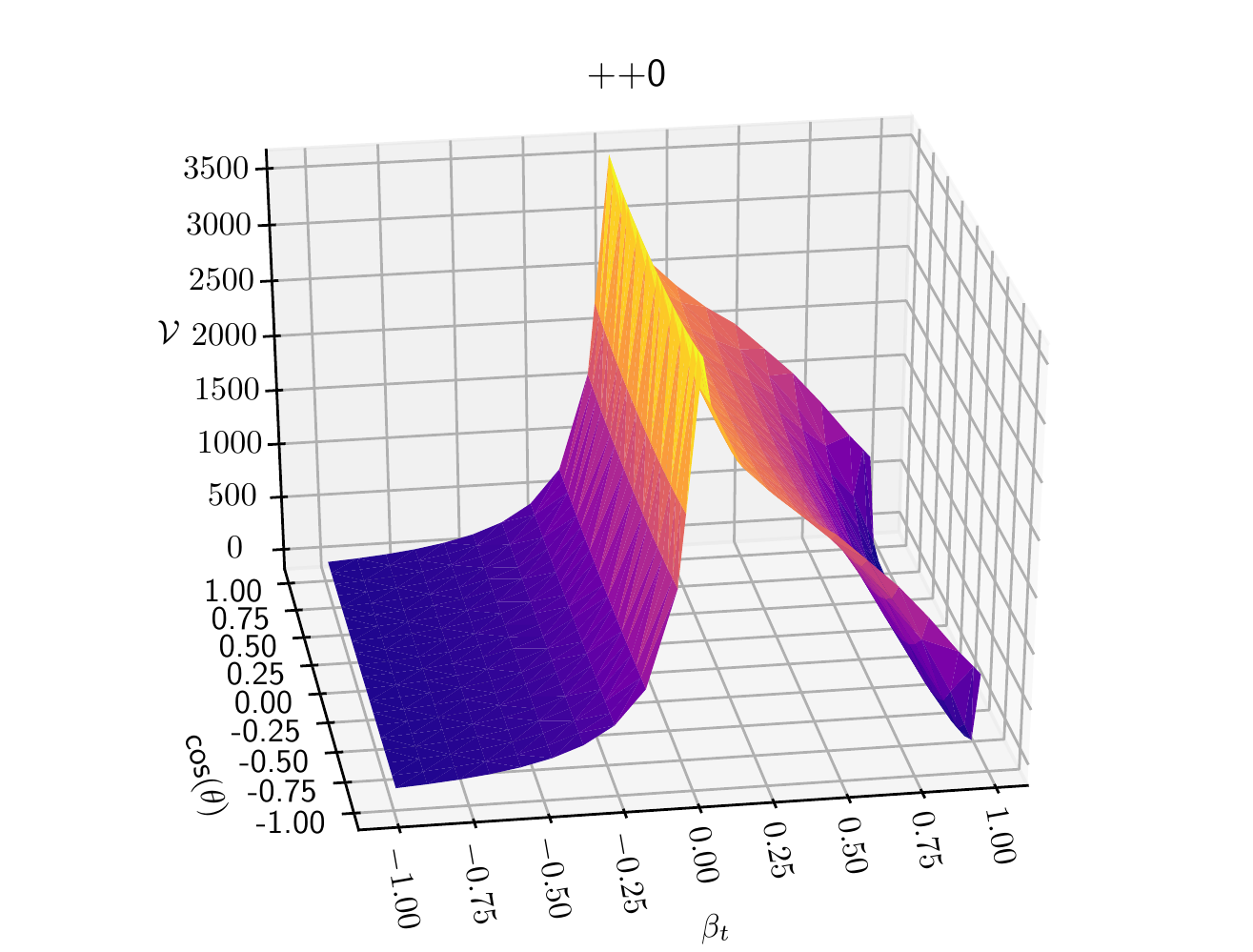}\\
\includegraphics[width=0.49\textwidth]{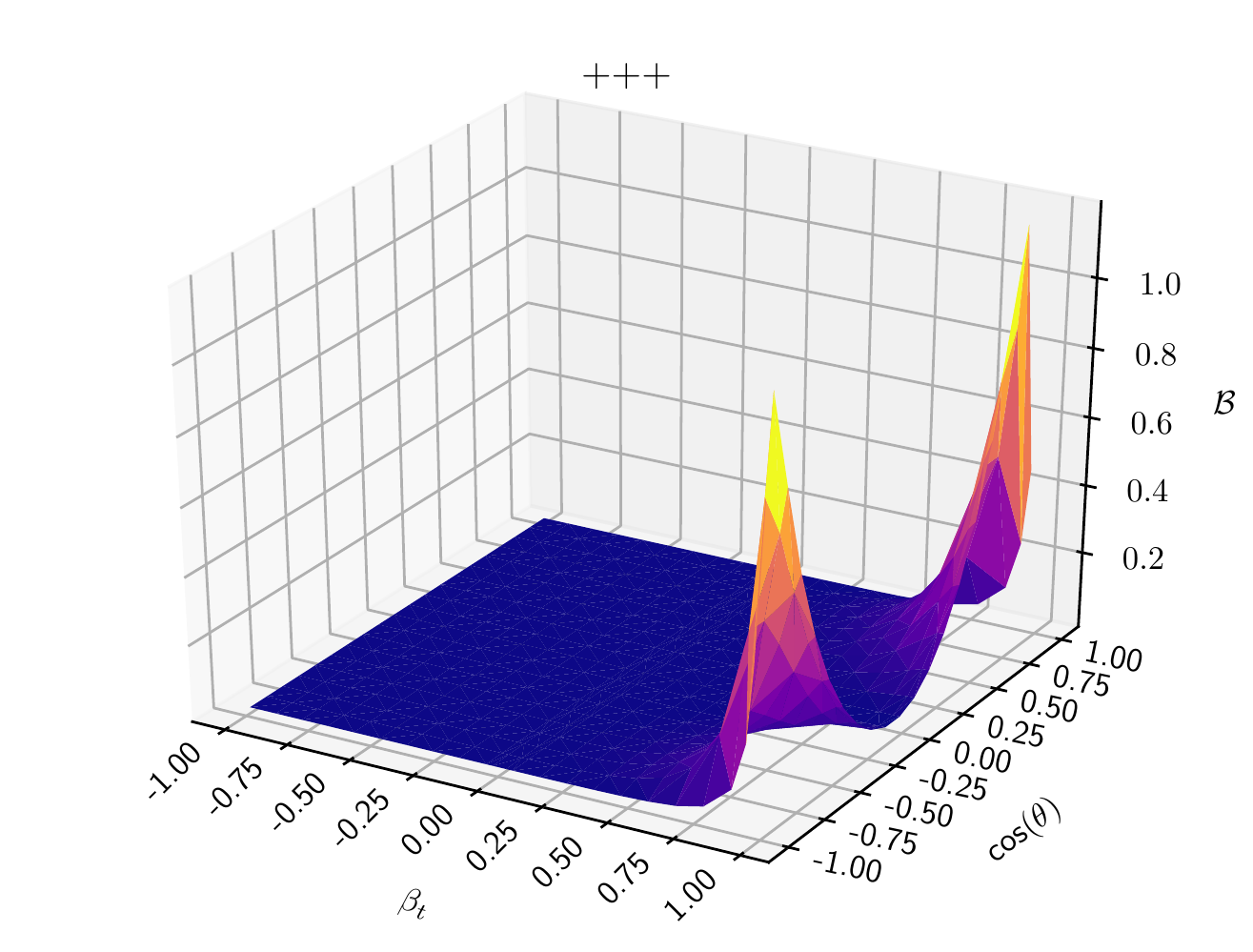}
\includegraphics[width=0.49\textwidth]{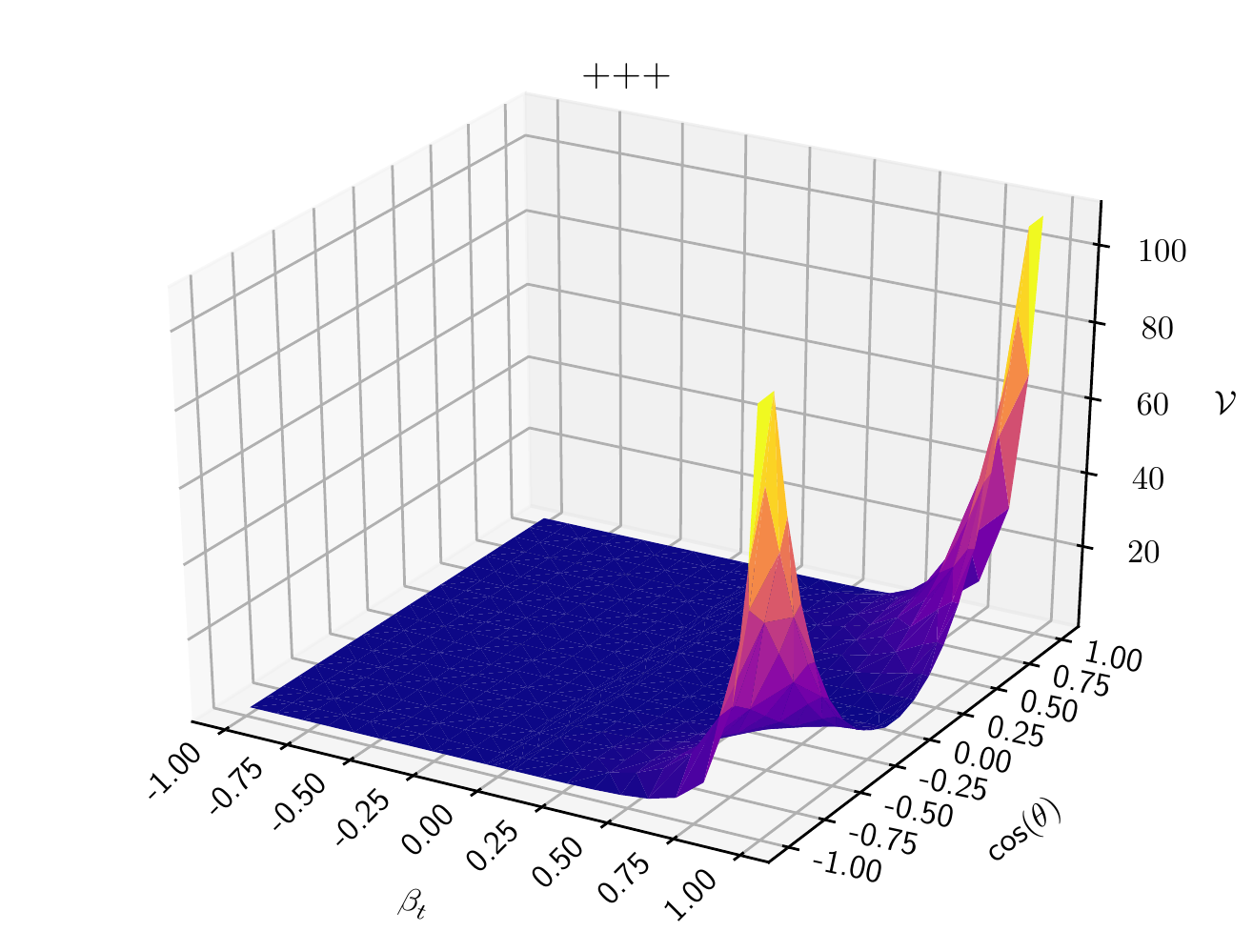}\\
\includegraphics[width=0.49\textwidth]{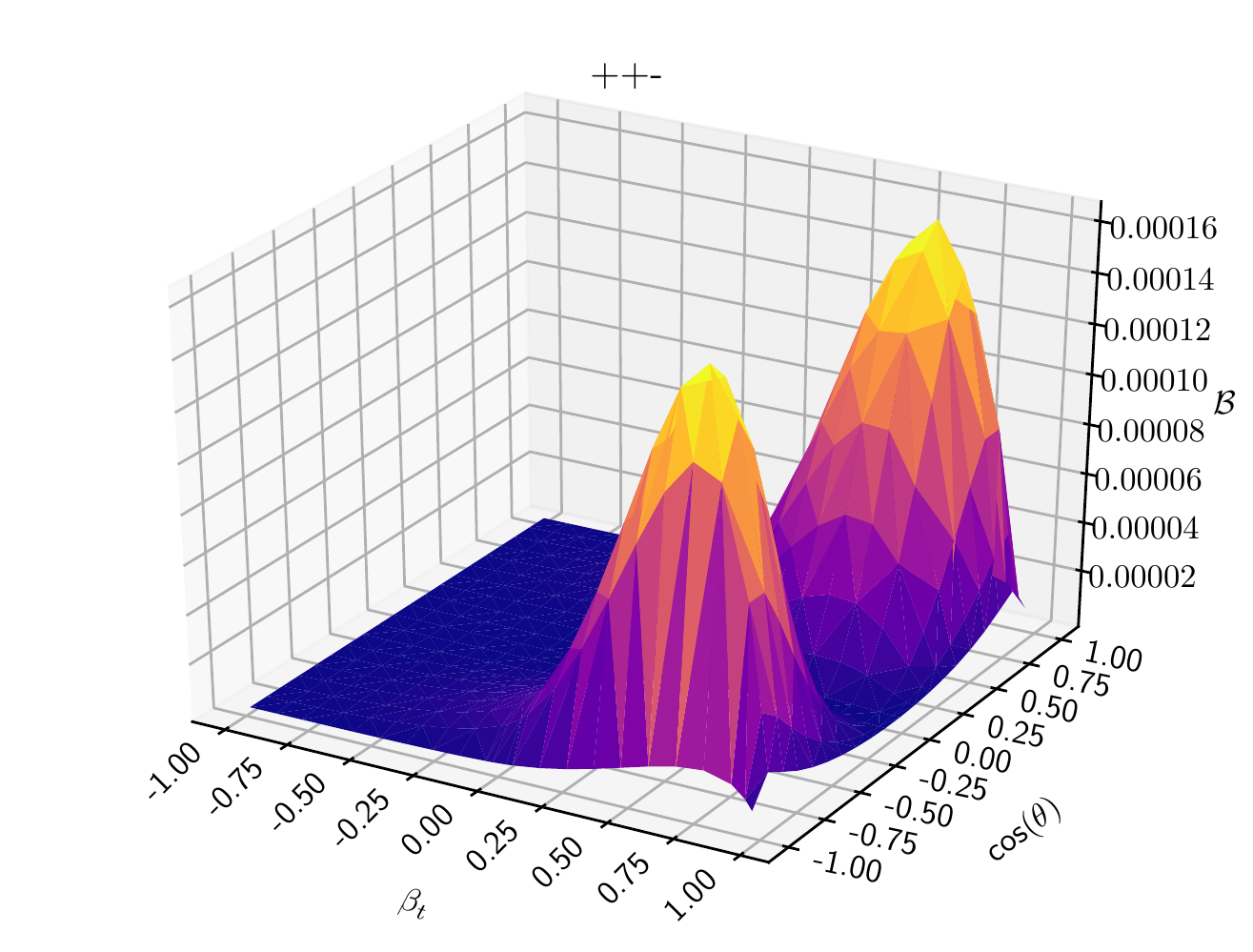}
\includegraphics[width=0.49\textwidth]{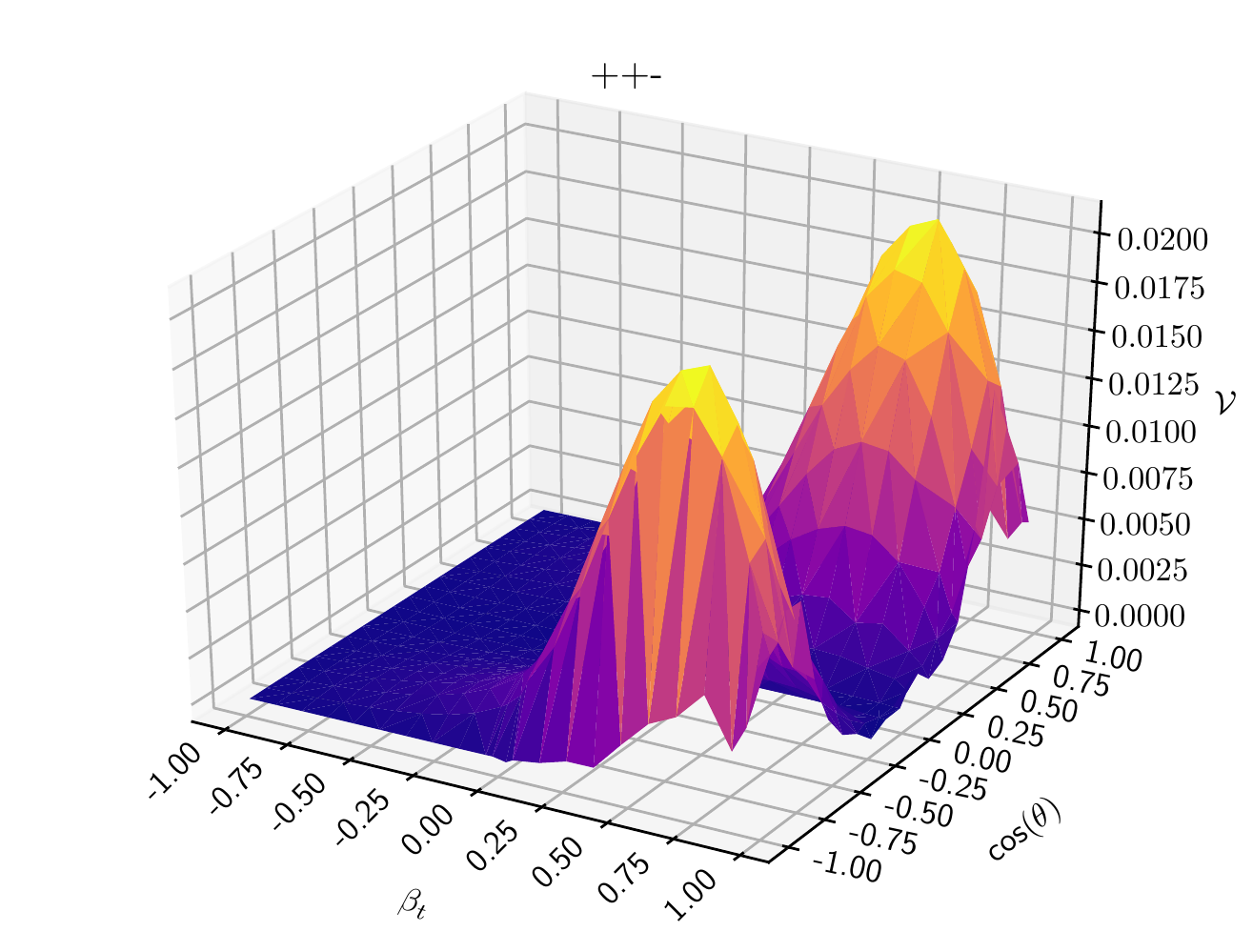}
  \caption{Dependence of the leading order~(left) and virtual
    contribution~(right) on the parameters $\beta_t$ and
    $\cos(\theta)$ for the individual helicity amplitudes with $s_i=0$.  }
\label{fig:helicityAmpsSpin0}
\end{figure}

From Fig.~\ref{fig:helicityAmpsSpin0}, we further observe that
the helicity amplitudes with the polarisations $\epsilon^Z_\pm $ are suppressed compared to those with  a longitudinally polarised $Z$-boson.
The amplitudes ${\cal A}_{++\pm}$ are antisymmetric under exchange of $t$ and $u$, so antisymmetric in $\cos\theta$.
The contributions with $J=1$, being proportional to $\sin\theta$, therefore do not occur in ${\cal A}_{++\pm}$.
The d-wave contributions $d_{0,\pm 1}^2(\theta)$ are proportional to $\pm \cos\theta\sin\theta$, which vanish at $\cos\theta=0, \pm 1$, a behaviour that can be observed in ${\cal A}_{++-}$. However, kinematics encoded in the coefficients of the partial waves also plays a major role, such that the shapes cannot be explained by partial waves alone.
Note that ${\cal A}_{++-}$ is about five orders of magnitude smaller
than ${\cal A}_{++0}$, and ${\cal A}_{+++}$ is also suppressed, therefore the amplitudes ${\cal A}_{++\pm}$ give a very minor contribution in the sum of all polarisation configurations.

\begin{figure}
\centering
\includegraphics[width=0.49\textwidth]{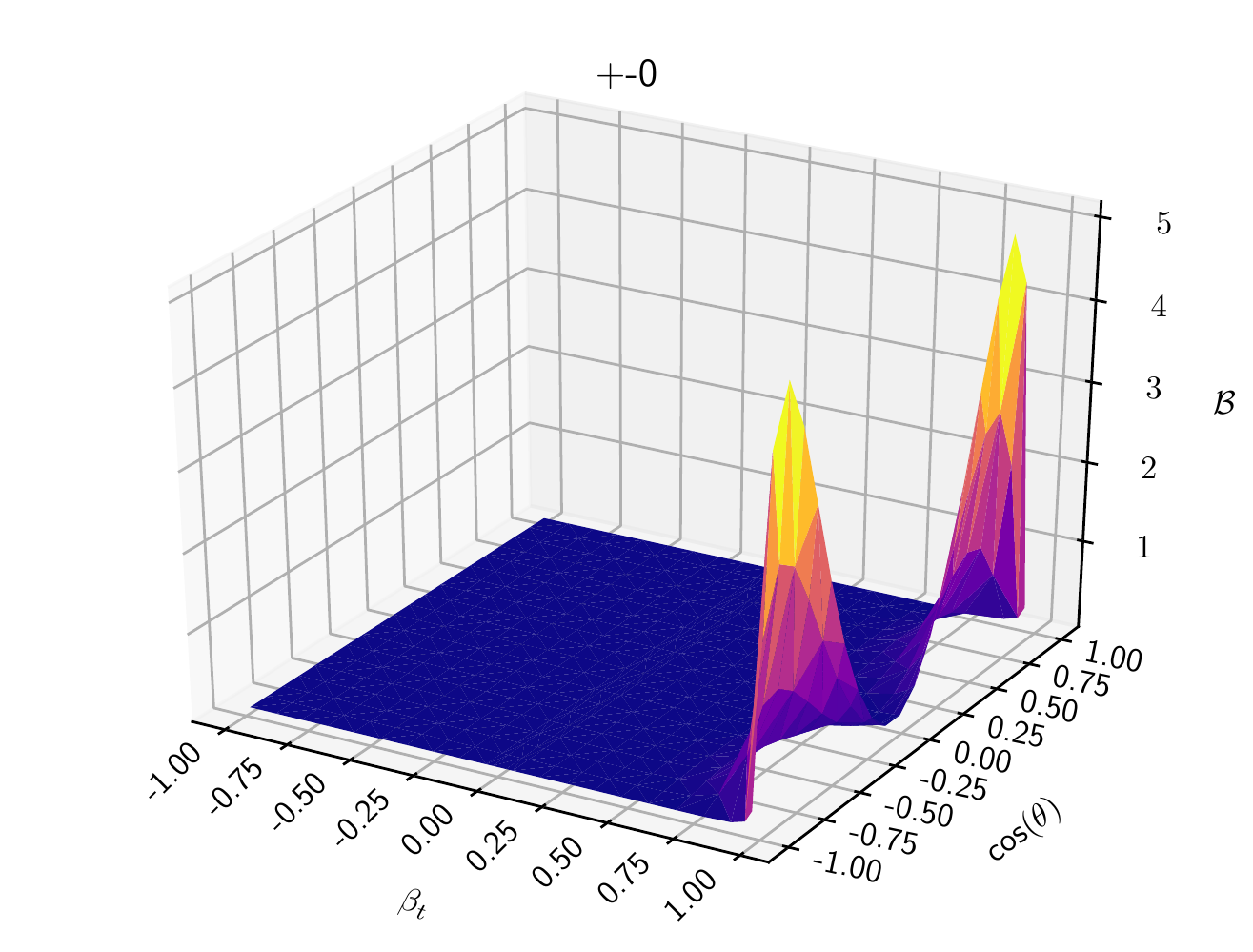}
\includegraphics[width=0.49\textwidth]{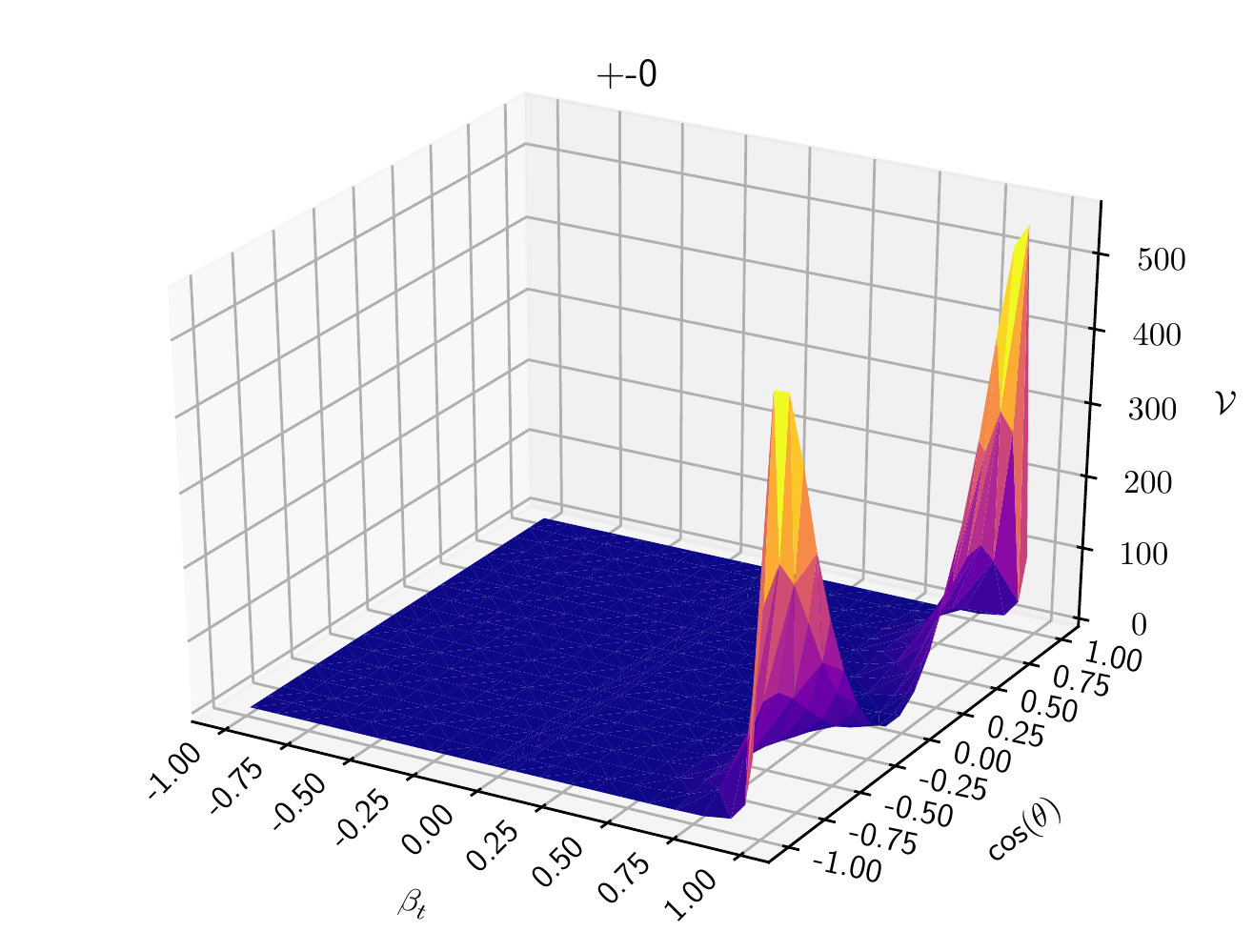}\\
\includegraphics[width=0.49\textwidth]{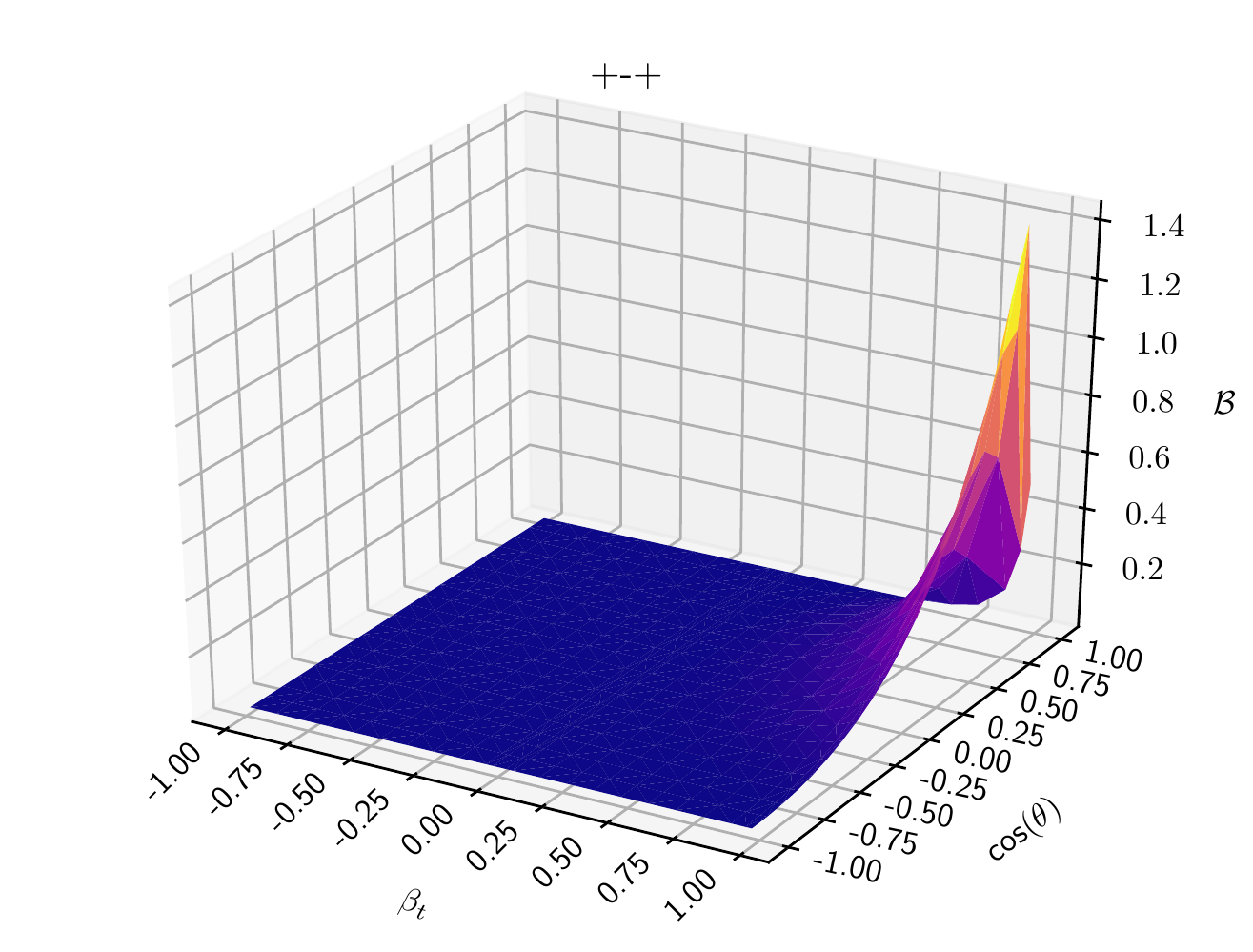}
\includegraphics[width=0.49\textwidth]{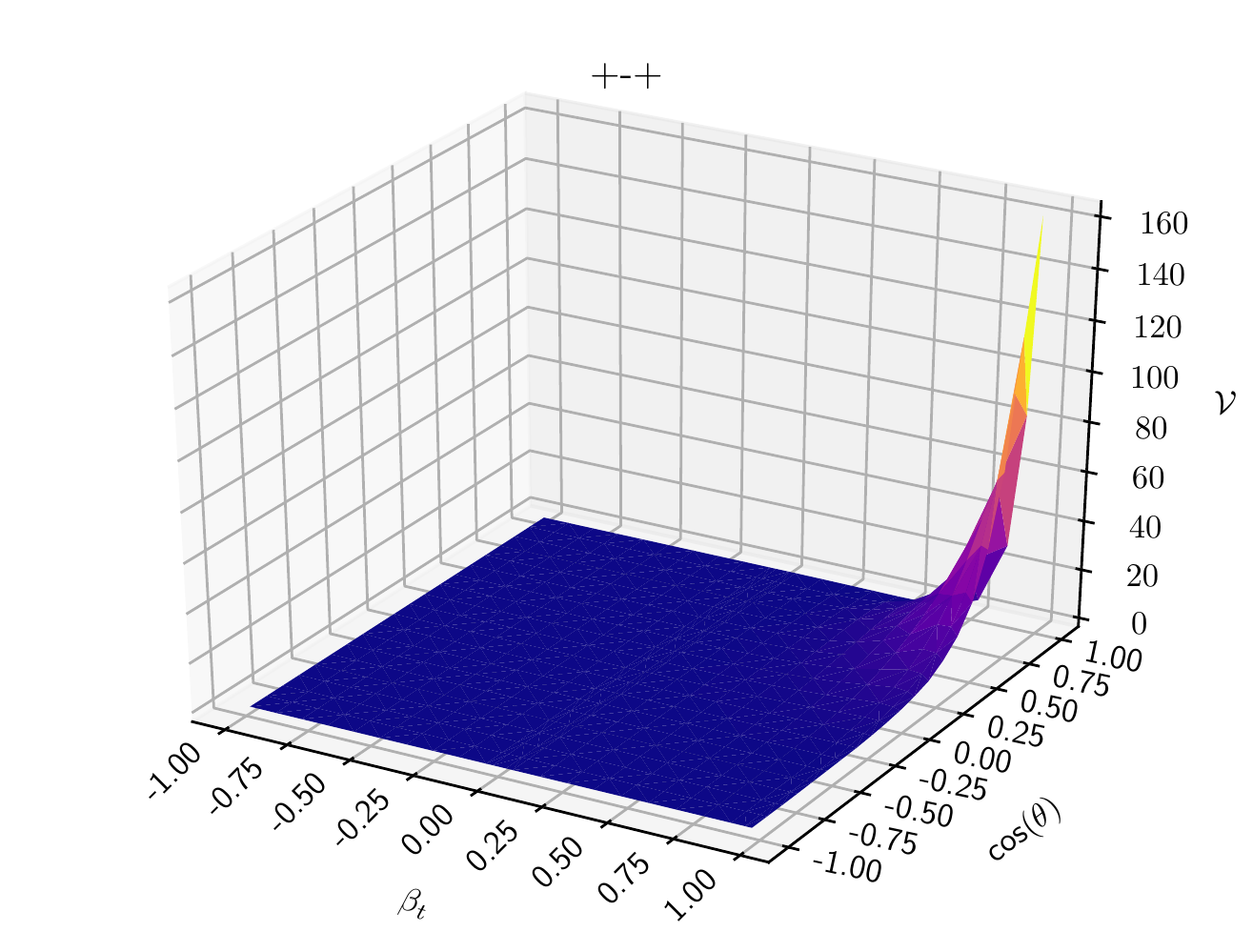}\\
  \caption{Dependence of the leading order~(left) and virtual
    contribution~(right) on the parameters $\beta_t$ and
    $\cos(\theta)$ for the individual helicity amplitudes with $s_i=2$.  }
\label{fig:helicityAmpsSpin2}
\end{figure}

For different helicities of the initial state gluons,
$(\lambda_1,\lambda_2)=(+,-)$ or $(-,+)$, shown in Fig.~\ref{fig:helicityAmpsSpin2},
the initial state has total spin $s_i=2$, such that the  partial wave contributions start from $J=2$. 
Therefore the amplitude ${\cal A}_{+-0}$ is much smaller than ${\cal A}_{++0}$.
This is also reflected in the fact that the amplitudes are basically zero except at very high energies.
The amplitude ${\cal A}_{+-0}$ has no contribution from
$d_{2,0}^J(\theta)$ with even $J$ as it is antisymmetric in $\cos\theta$,
therefore its leading partial wave is given by $d_{2,0}^3(\theta) \sim \cos\theta \sin^2\theta$, vanishing at $\cos\theta = \pm 1$ and $\cos\theta = 0$.
The amplitudes ${\cal A}_{+-\pm}$ have their leading partial waves given by $d_{2,\pm 1}^2(\theta)\sim \sin\theta(1\pm\cos\theta)$.
Consequently ${\cal A}_{+-+}$ is highly suppressed in the backward direction, while ${\cal A}_{+--}$ is highly suppressed in the forward direction.

In Table~\ref{tab:amplitudeResults} we list numerical results of the amplitude to facilitate comparisons with our results.

  \begin{table}
    \centering
\begin{tabular}{rr|rr}
    $s/m_t^2$ & $t/m_t^2$  & $\mathcal B$ & $\mathcal V$ \\ \hline
1.707133657190554 &  -0.441203767016323 &   0.4412287 &   35.429092(6) \\
3.876056604162662 &  -1.616287256345735 &  37.2496999 &   4339.045(1)  \\
4.130574250302561 &  -1.750372271104745 &  66.3224970 &   6912.361(3)  \\
4.130574250302561 &  -2.595461551488002 &  67.1908198 &   6981.09(2)  \\
134.5142052093564 &  -70.34125943305149 &   4.1920928 &   -153.9(4)  \\
134.5142052093564 &  -105.1770655376327 &  14.7405104 &   527(4)
\end{tabular}
  \caption{Numerical results for various phase space points at the scale $\mu_R^2 = s$.
    The number in parentheses gives the numerical uncertainty on the last digit of the virtual amplitude.}
\label{tab:amplitudeResults}
\end{table}

\section{Conclusions and Outlook}
\label{sec:conclusions}

We have numerically calculated the two-loop amplitudes for the production of a Higgs- and a $Z$-boson in gluon fusion with massive top quark loops.
The results for the finite part of the two-loop amplitude interfered with the Born amplitude
are plotted as a function of the scattering angle and the centre-of-mass energy, for the total unpolarised amplitude as well as for individual helicity amplitudes.

The projection of the amplitudes to scalar quantities has been carried out with projectors onto linear polarisation states, from which helicity amplitudes are constructed~\cite{Chen:2019wyb}.
The reduction to master integrals has been performed with the program \texttt{Kira}~\cite{Maierhoefer:2017hyi,Klappert:2020nbg} in combination with the rational function interpolation library \texttt{FireFly}~\cite{Klappert:2019emp,Klappert:2020aqs}, using in addition \texttt{LiteRed}~\cite{Lee:2013mka} and \texttt{Reduze}~\cite{vonManteuffel:2012np} to obtain dimensional recurrence relations.
The master integrals have been calculated using \pysecdec~\cite{Borowka:2017idc,Borowka:2018goh}.
The integration is sufficiently stable and accurate, also in the near-threshold and forward scattering regions,
to perform phenomenological studies based on these results, after including the real radiation contributions.
We postpone such a phenomenological analysis to a subsequent publication.

Our method for the first time has been applied to a process with three different mass scales, $\mt, \mh$ and $\mz$.
We expect that it can be applied successfully  to other two-loop amplitudes involving several mass scales in the future.

\section*{Acknowledgements}
We would like to thank Joshua Davies, Go Mishima and Matthias
Steinhauser for the comparison of results prior to publication.
We also would like to thank Tom Zirke for conversation about the
large mass expansion and Ramona Gr\"ober for useful discussions.
This research was supported by the Deutsche Forschungsgemeinschaft (DFG, German Research Foundation) under grant  396021762 - TRR 257,
by the COST Action CA16201 (`Particleface') of the European Union 
and by the Swiss National Science Foundation (SNF) under grant number 200020-175595.
MK acknowledges support by  the Forschungskredit of the University of Zurich, grant no.  FK-19-102. 
The research of JS was supported by the European Union through the ERC Advanced Grant MC@NNLO (340983).
SJ is supported by a Royal Society University Research Fellowship (Grant URF/R1/201268).
We gratefully acknowledge resources provided by the Max Planck Computing and Data Facility (MPCDF).
Some of the calculations were performed with computing resources granted by RWTH Aachen University under project rwth0541.

\bibliographystyle{JHEP}
\bibliography{refs_zh}

\end{document}